\documentclass[10pt,a4paper]{article}
\usepackage[utf8]{inputenc}
\usepackage{lipsum}
\usepackage{amsmath}
\usepackage{mathtools}
\usepackage{amsfonts}
\usepackage{amssymb}
\usepackage{amsbsy}
\usepackage{float} 
\usepackage{graphicx}
\usepackage{cite}
\usepackage{caption}
\usepackage{subcaption}
\usepackage{multicol}
\usepackage[T1]{fontenc}
\usepackage[a4paper,left=1in,right=1in,top=1in]{geometry}
\begin{document}
\pagestyle{empty}
\begin{flushright}
\textit{(Statistical Mechanics of Soft Condensed Matter Physics)}
\end{flushright}
\begin{center}
\Large{\textbf{Current and diffusion of Overdamped Active Brownian Particles in a Ratchet Potential}}\\[.1in]
\normalsize {{{ Arjun S R$ ^{1} $} , Ronald Benjamin$ ^{1} $}\\[.1in]
\noindent
\textit{$ ^{1} $Department of Physics, Cochin University of Science and Technology, Cochin-22, Kerala ,India}}\\
\small { {rbenjamin@cusat.ac.in, arjunsr@cusat.ac.in}}\\
\end{center}
\noindent
\rule{\textwidth}{0.4pt}\\[-0.2cm]

\textbf{\large{Abstract}}\\[0.05in]
The transport properties of a spherical active brownian particle in a periodic potential under heavy damping is considerd. The self propelled particle is subjected to the asymmetric potential, detailed balance is lost and the particles generate a non zero drift speed. The average current is calculated and diffusivity of the particle is analysed from the effective diffusion coefficient. For chiral active particles the diffusivity decreases with increasing the angular velocity, confining the particle near the initial position, reducing the average current.    \\
\noindent
\rule{\textwidth}{0.4pt}
\begin{multicols}{2}

\section{Introduction}
Brownian or molecular motors is an active area of research where the main focus is to obtain rectifying motion in the Brownian regime where the Brownian particles are driven by non-equilibrium fluctuations of zero mean. In the macroscopic domain, when considering the external forces, the thermal force is almost negligible. But to confine particles and to obtain work done in the Brownian regime, energy has to be spent constantly. There comes the importance of the ratchets where the external force on an average is zero, but useful work is derived\cite{reichhardt2016ratchet}.\par

The Active Brownian Particles(ABP) are different from the passive
by performing an active motion composed by an internal energy depot(depot model)\cite{ebeling1999active} or/and a friction function, dependent on velocity. The ABP were analysed with the help of simple dynamic models derived from the langevin equations and study the dynamics of individual ABP. The analytical, numerical and computational methods are integrated to study the dynamics and properties of such complex sytems.

The extraction of work from the Brownian particles subjected to thermal noise can be classified as a simple model of Feynman Ratchet\cite{reimann2002brownian}. Considering such a system, it is clear that only diffusion occurs in the presence of just thermal noise, and no effective transport will be obtained\cite{volpe2013simulation}. The conditions for a Brownian particle to exhibit directional transport are broken spatial symmetry and an external force with time correlations\cite{magnasco1993forced}. In the case of ABPs they are manifestly out of equilibrium and the asymmetry in the potential generates a current and it's direction depends on the asymmetric parameter. The self proppeled velocity helps the particle to overcome the potential towards the gentle side and the steeper side restricts it from moving backwards generating a dc current with out any external source.  \par

The paper is divided into four sections. First the analytical formulation with the Fokker- Planck equation. Then focuses on the dynamics of an overdamped active Brownian particle in an asymmetric ratchet potential. The various quantities of interest are the current, effective diffusion coefficient, position correlation, probability distributions, and the peclet number. The third section focuses the energetics which includes efficiency(stocks efficiency, thermodynamic efficiency, and generalized efficiency), entropy, heat and energy input. And the last section is the conclusion.
 
 \section{Basic Model and Methods}
 	We are considering one dimensional straight channel where the particles experience a periodic potential $U(x) = U(x+L)$ with a period $L$. The potential can defined with an asymmetric parameter $\alpha$ which can take values $0<\alpha<1$.\par
 	\begin{equation}
 	U(x)=\begin{cases}
 	u_0\frac{x}{\alpha}, &0\le x<\alpha\\
 	u_0\frac{L-x}{L-\alpha},&\alpha<x<L
 	\end{cases}
 	\end{equation}
 	where, $u_0$ is the height of the potential.\par  
 	 A single spherical Active Brownian particle is considerd whose dynamics are predicted by the position, descibed by $x$ of its center and the polar axis orientation $\theta$ . The Langevin equation\cite{ai2017transport} can be written as, 
 	  \begin{align}
 	 \gamma_x \frac{dx}{dt}&=v_0cos(\phi)+ f(x)+\xi_x(t)\\
 	 \gamma_{\phi} \frac{d\phi}{dt}&=\Omega+\xi_{\phi}(t)
 	 \end{align}
 	 where $v_0$ is the self proppelled velocity of the Active Brownian Particle, $f(x)$ is the force acting on the particle from the substrate, $\gamma_x$ and $\gamma_{\phi}$ are the friction coefficients having the values $6\pi \eta R$ and $8 \pi \eta R^3$ respectively where $\eta$ is the viscocity of the medium and $R$ is the radius of the particle. $\Omega$ is the angular velocity which is produced as an output of the torque acting on the ABP\cite{volpe2014simulation}. The sign of the $\Omega$ describes whether the ABP bends in clockwise or anticlockwise direction.\par
 	\begin{figure}[H]
 		\centering
 		\includegraphics[height=2in]{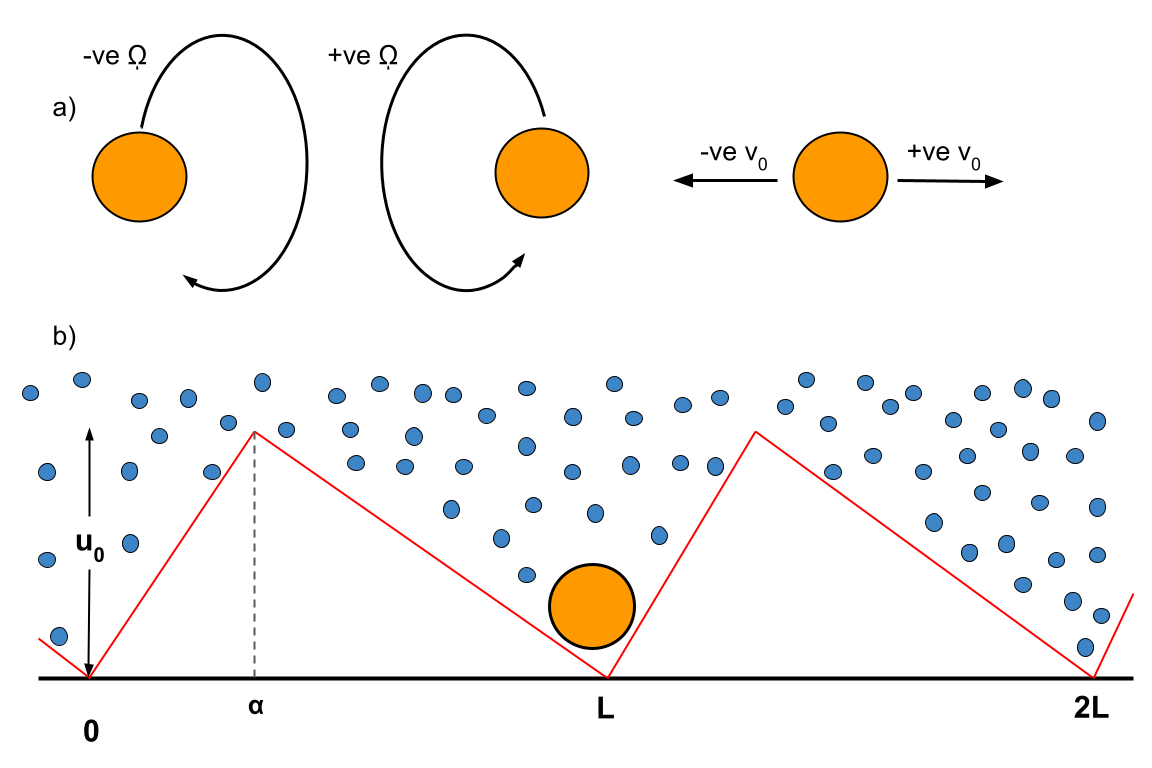}
 		\caption{Active Brownian Particle in Piecewise linear potential}
 	\end{figure}

 The average current generated by the ABP's when they attains stationary state can be calculated from the Fokker-Planck equation\cite{sevilla2015smoluchowski}.
  \begin{align}
  \MoveEqLeft[3]  \frac{\partial P(x,\phi,t)}{\partial t} + v_0cos(\phi)\frac{\partial P(x,\phi,t)}{\partial x} \notag\\ 
  ={}& \frac{\partial (P(x,\phi,t)U'(x))}{\partial x}+D_T \frac{\partial^2 P(x,\phi,t)}{\partial x^2}\notag\\
 & +D_R \frac{\partial^2 P(x,\phi,t)}{\partial \phi^2}
  \end{align}
 For very small values of $D_R$ the last term can be neglected. At stationary state the average current generated by the ABP,s will be constant. From the equation of continuity, we can write,
   \begin{equation}
  J= (-u'(x) + v_0cos(\phi))P(x,\phi)-D_T \frac{\partial P(x,\phi)}{\partial x}
  \end{equation}
  Since $U'(x)$ have two values for $0 \le x < \alpha$ and $\alpha \le x \le 1$, we can write
  \begin{alignat}{1}
    J_1 &=(-\frac{u_0}{\alpha} + v_0cos(\phi))P(x,\phi)\notag\\
    & -D_T \frac{\partial P(x,\phi)}{\partial x},\;\;\;\;\; 0 \le x \le \alpha\\
  J_2 &=(\frac{u_0}{1-\alpha} + v_0cos(\phi))P(x,\phi)\notag\\
    & -D_T \frac{\partial P(x,\phi)}{\partial x},\;\;\;\;\;\alpha \le x \le 1
  \end{alignat}
  
 Solving these equations and considering the fact that there will be no source and sink, so $J_1=J_2=J$ and applying the boundary conditions,
  \begin{align}
  P_1(0)&= P_2(1)\\
   P_1(\alpha)&= P_2(\alpha)\\
   \int_{0}^{\alpha}P_1(x)dx+&\int_{\alpha}^{1}P_2(x)dx=1
  \end{align}
  Finally the expression for the current can be written as
   \begin{equation}
  J(x,\phi)= \frac{mr-p}{q(mr-p)-a(mn+h)}
  \end{equation}
  where,
   \begin{align}
  a&= \frac{1}{\beta_1} - \frac{1}{\beta_2}\\
  b&= exp\biggl(\frac{\beta_2}{D_T}\biggr)\\
  r&=  exp \biggl(\frac{\beta_1\alpha}{D_T}\biggr)\\
  p&=  exp\biggl(\frac{\beta_2\alpha}{D_T}\biggr)\\
  q&=\frac{\alpha}{\beta_1} + \frac{1}{\beta_2}-\frac{\alpha}{\beta_2}\\
  n&= \frac{D_T}{\beta_1}\Biggl[exp\biggl(\frac{\beta_1\alpha}{D_T}\biggr) - 1\Biggr]\\
  h&= \frac{D_T}{\beta_2}\Biggl[exp\biggl(\frac{\beta_2}{D_T}\biggr) - exp\biggl(\frac{\beta_2\alpha}{D_T}\biggr)\Biggr]\\
  m&=\frac{p-b}{r-1}
  \end{align}
  The potential height, and selfpropelled velocities are incorperated in $\beta_1$ and $\beta_2$
  \begin{align}
  \beta_1&= \frac{-u_0}{\alpha} + v_0cos(\phi)\\
  \beta_2&= \frac{u_0}{1-\alpha} + v_0cos(\phi)
  \end{align}

The solution gives the value of current as a function of $x$ and $\phi$. Finally the current can be optained by averaging the it over all possible values of $\phi$.
\begin{equation}
J_{av}(x)=\int_{0}^{2\pi}J(x.\phi)d\phi
\end{equation}
It will be a bit complicated to integrate the final equation. We can apply numerical integration to obtain the solution and can compare with the simulation results.
   \section{Results and Discussions}
  The dynamics of the particle is studied by analysing the current and effective diffusion coefficient. The value of effective diffusion coefficient gives an idea about the diffusivity of the particle and how the diffusivity changes with other parameters.
   The parameters considering here are the self-proppeled velocity of the active browninan particle, it's angular velocity, Translational diffusion coefficient, rotational diffusion coefficient, height of the potential and the asymmetric parameter.\par 
   When the asymmetric parameter is varied a current reversal is observed and the average current will be zero when the potential is symmetric. The  selfproppeled velocity allows the particle to move towards the gentle side of the potential(Figure 2).\par
    \begin{figure}[H]
   	\includegraphics[scale=0.047]{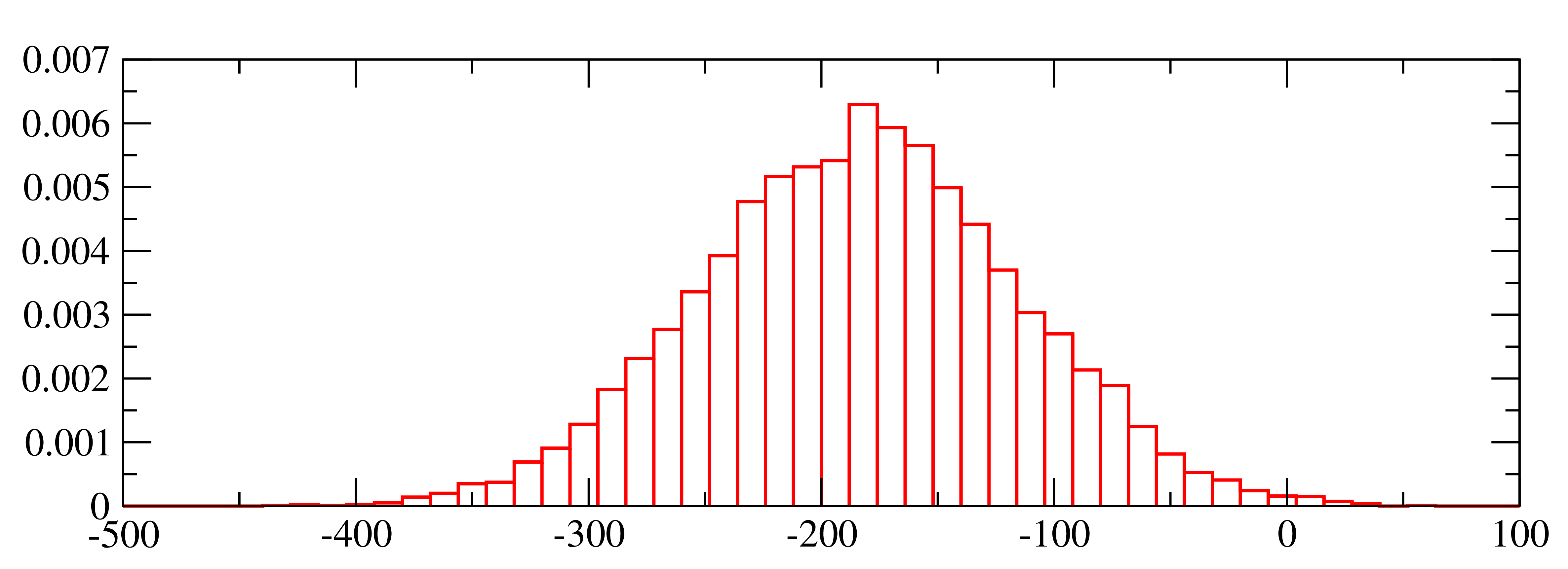}
   	\caption{Probability distribution of the particle when they reach steady state. The value of $\alpha$ is 0.09 generating current towards negative x direction.} 
   \end{figure}
  When the current is plotted as a function of asymmetric parameter, the current decreases with increase in value of $D_T$ and reaches 0 for very large values of $D_T$ and the diffusivity increases with increase in $D_T$(Figure 3). The increase in value of $D_T$ decreases the current and increases the diffusivity indicates that the for larger values of $D_T$ the particle overcome the asymmetry of the potential in both directions generating 0 current but transports far away from the initial position. It is also observed that for large values of $D_T$ the diffusivity increases when the asymmetry is reduced and shows maximum diffusivity when the potential is symmetric and vice versa. The average current increases with increasing values of the self proppelled velocity and there is also an increase in the effective diffusion coefficient(Figure 4).\par
  The effective diffusion shows a consistent behaviour with the increase in $D_R$. In Figure(5) when the $D_R$ increases the diffusivity decreases and for very small values of $D_R$ maximum diffusivity is observed. The current shows an increase with increasing values of $D_R$(Figure 6).\par
  The current increases with increasing values of $D_T$ and reaches a maximum. There is a critical value of $D_T$ after that the current decreases. The diffusivity increases with increasing values of $D_T$(Figure 7). With increase in $D_R$ the current and effective diffusion coefficient decreases. With increase in $v_0$ the current increases but effective diffusion coefficient decreases(Figure 8) \par
  The behaviour of chiral active Brownian particles is an intresting area to explore. The active matter rotates with an angular velocity $\Omega$. It is due to the torque acting on the particle\cite{kummel2013circular}. The chirality swimming patterns is observed in many micro-organisms like ecoli bacterias, salomnellas in the the surface of the water\cite{elgeti2015physics}. From the Figure(9) its observed that chirality reduces the current. Chirality confines the particle near the initial position. It is also observed that the direction of the angular momentum does not hinders the transport properties of the particle, since it is symmetric for negative and positive values. The chirality doesnt allow the particle to ovecome the potential in the steeper side.\par  
  The increasing in selfproppeled velocity also increases the current and the effective diffusion coefficient(Figure 10).\par
  Plotting the current and effective diffusion coefficient as a function of potential height shows the trapping of the particle in the potential. If we increase the velocity the inially the current will increase. For higher potential heights the current will be zero(Figure 11). The same behaviour is observed for effective diffusion coefficient. With increasing values of $D_R$ the current and effective diffusion coefficient are reduced(Figure 12)\par   
  The average current and effective diffusion coefficient shows a symmetric behaviour when plotted as a function of self-proppelled velocity. Like the initial direction of angular velocity, the inital direction of the self propelled velocity also does not hinders the diection of transport. The direction of transport is completly controlled by the asymmetry of the ratchet. The magnitudes of current and diffusivity can only be altered by other parameters like self-proppelled velocity, barrier height, chirality etc.. The velocity decreases with increasing values of $D_T$ and $u_0$. But the effective diffusion shows a different behaviour. The diffusivity increases with increasing value of $D_T$(Figure 13) and decreases with increasing value of $u_0$(Figure 14). The barrier height reduces the current and increases the diffusivity.  
       
\end{multicols}
\newpage

  \begin{figure}[H]
  	\includegraphics[scale=0.045]{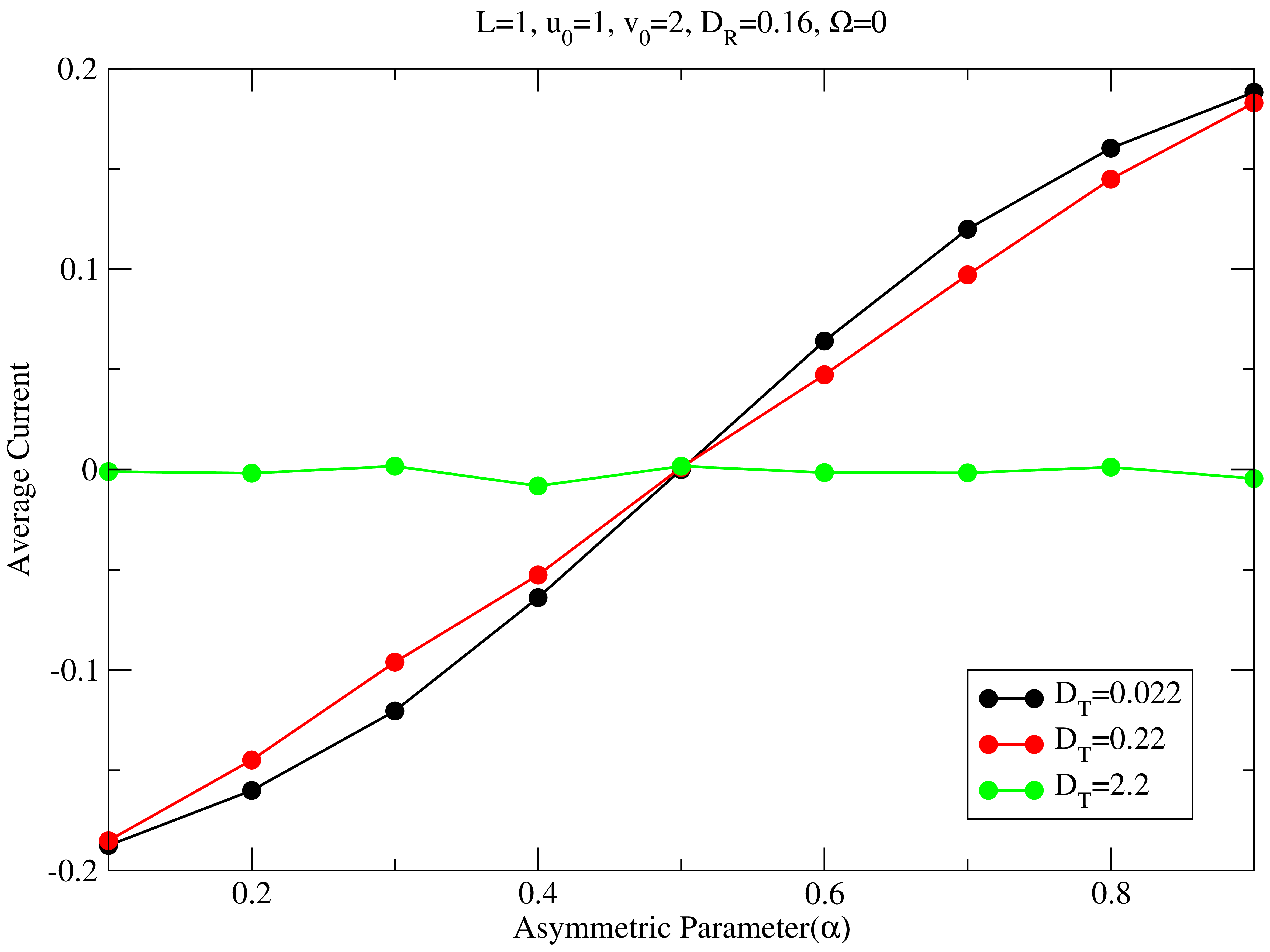}
  	\includegraphics[scale=0.045]{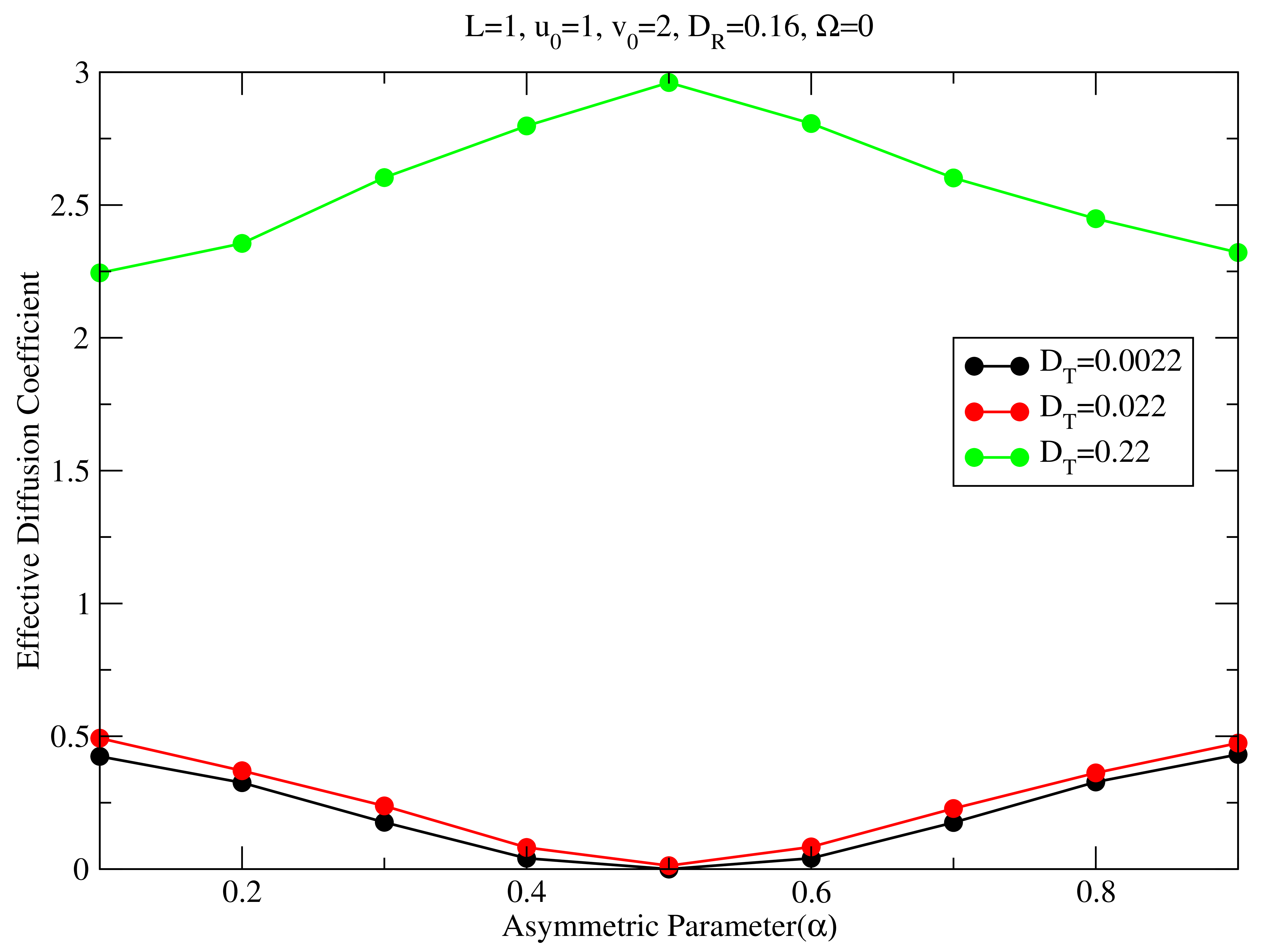}
  	 \caption{Average Current and Effective Diffusion Coefficient as a function of asymmetric parameter of an achiral particle for various values of $D_T$. The period of the potential is 1, height of the potential is 1, self proppelled velocity is 2, Rotational diffusion coefficient is 0.16, }
  \end{figure}

   \begin{figure}[H]
   	\includegraphics[scale=0.045]{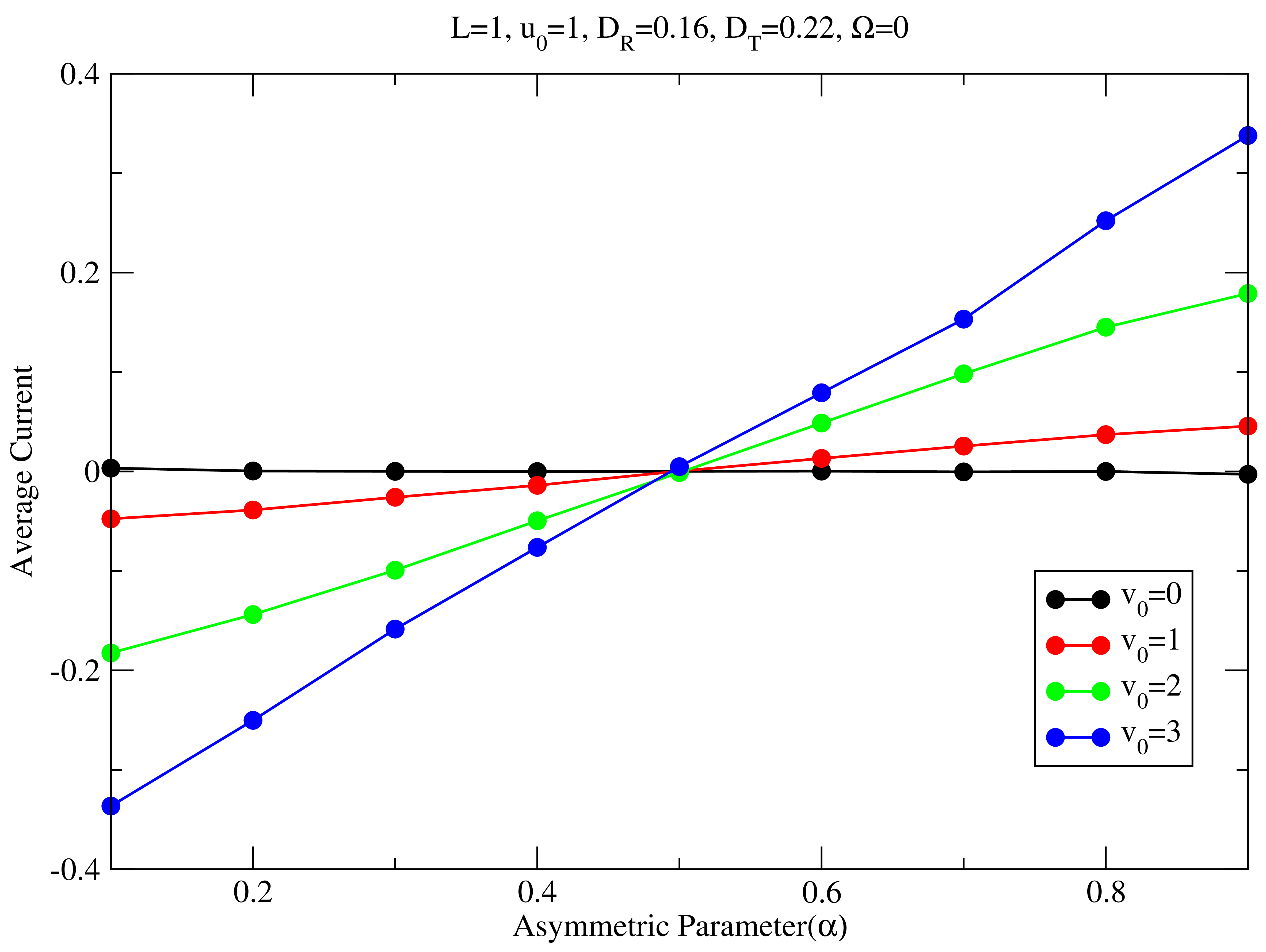}
   	\includegraphics[scale=0.045]{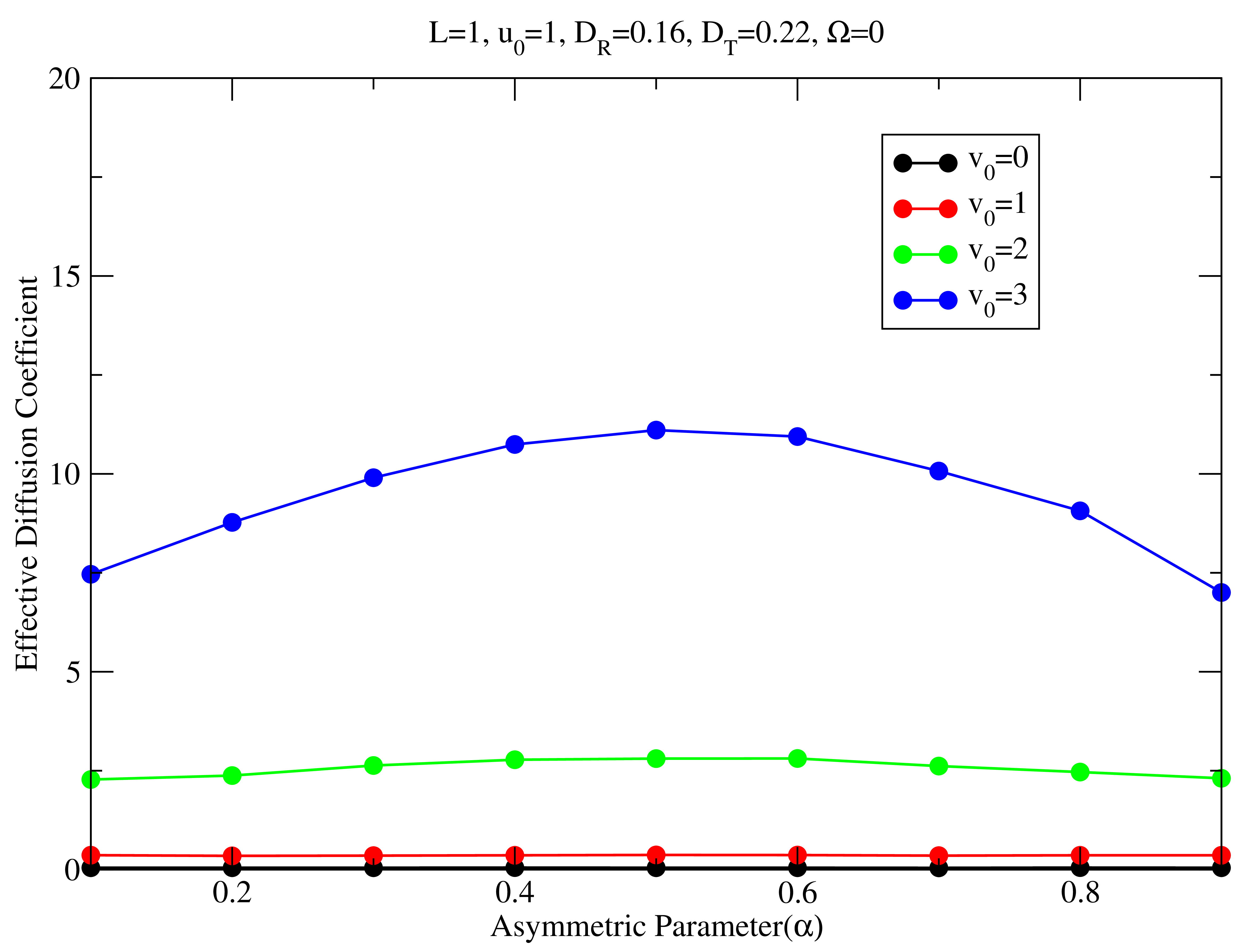}
   	\caption{Average Current and effective diffusion coefficient as a function of asymmetric parameter of an achiral particle for various values of $v_0$. The period and height of the potential is taken as unity, Translational diffusion coefficient is 0.22 and the Rotational diffusion coefficient is 0.16}
  	\end{figure}

  \begin{figure}[H]
  \includegraphics[scale=0.045]{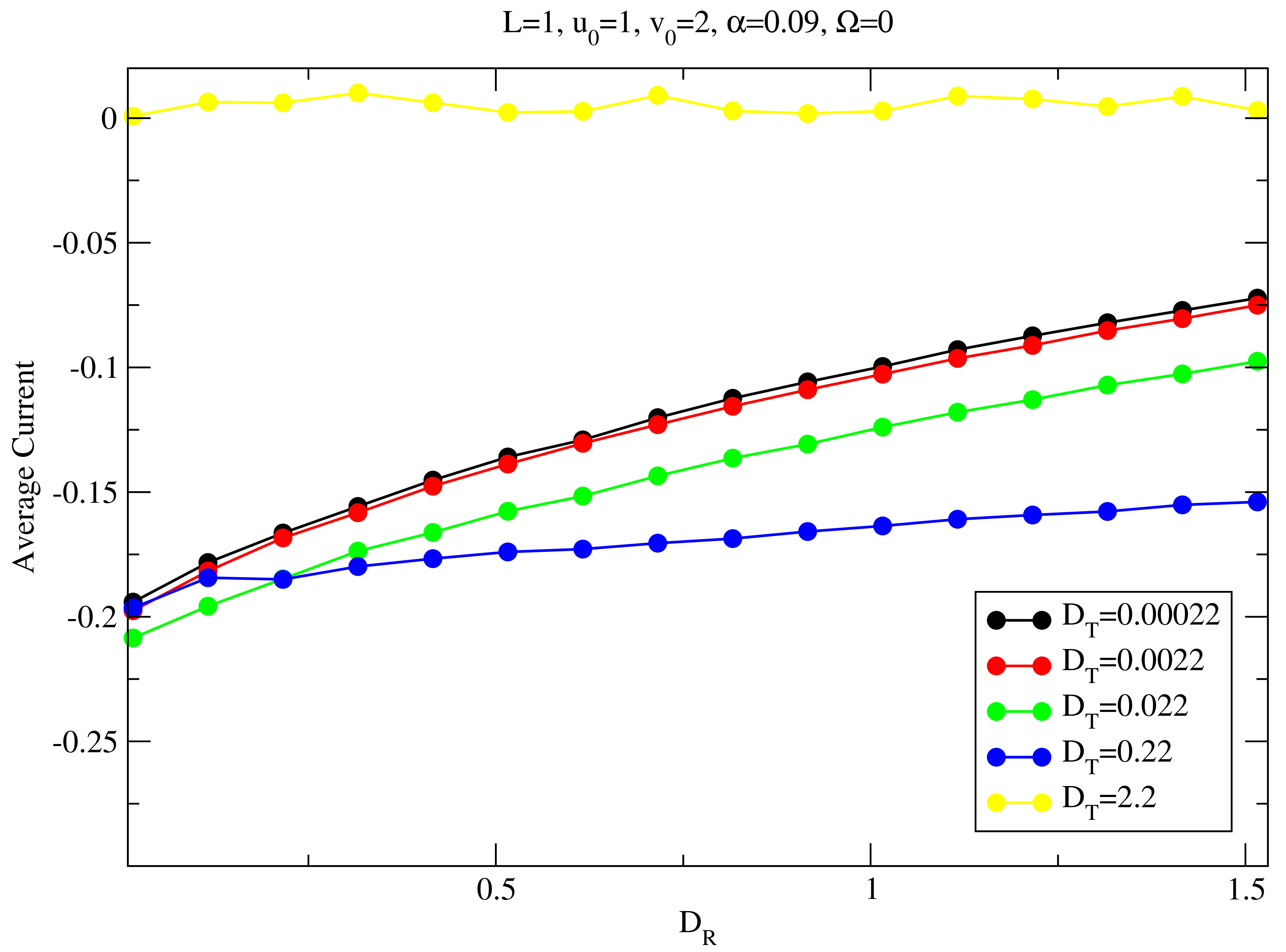}
  \includegraphics[scale=0.045]{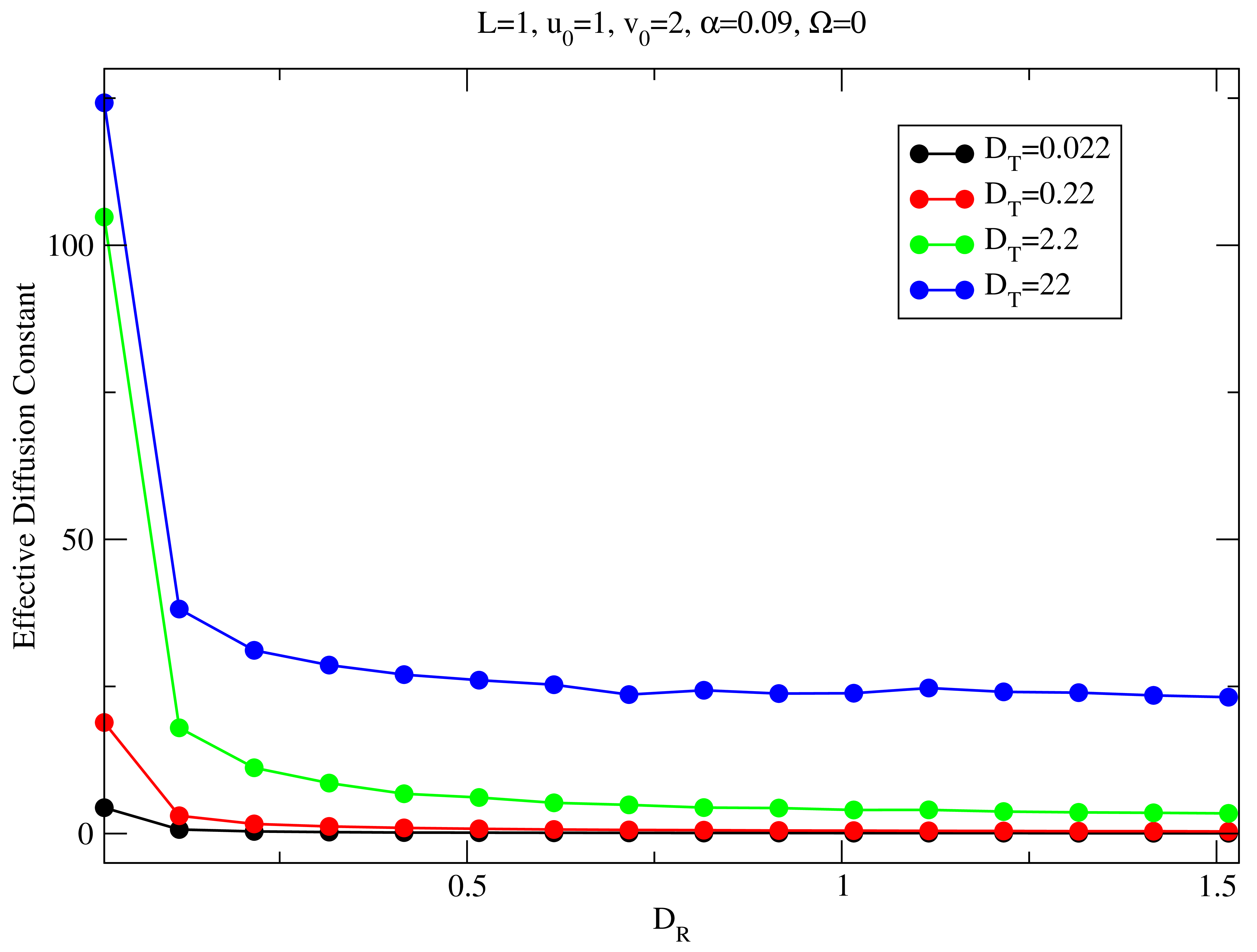}
  \caption{Average Current and Effective diffusion coefficient as a function of $D_R$ of an achiral particle for various values of $D_T$. The period and height of the potential is taken as unity, The self proppelled velocity is 2, and the value of asymmetric parameter is 0.09} 
   \end{figure}

\newpage
  \begin{figure}[H]
  \includegraphics[scale=0.045]{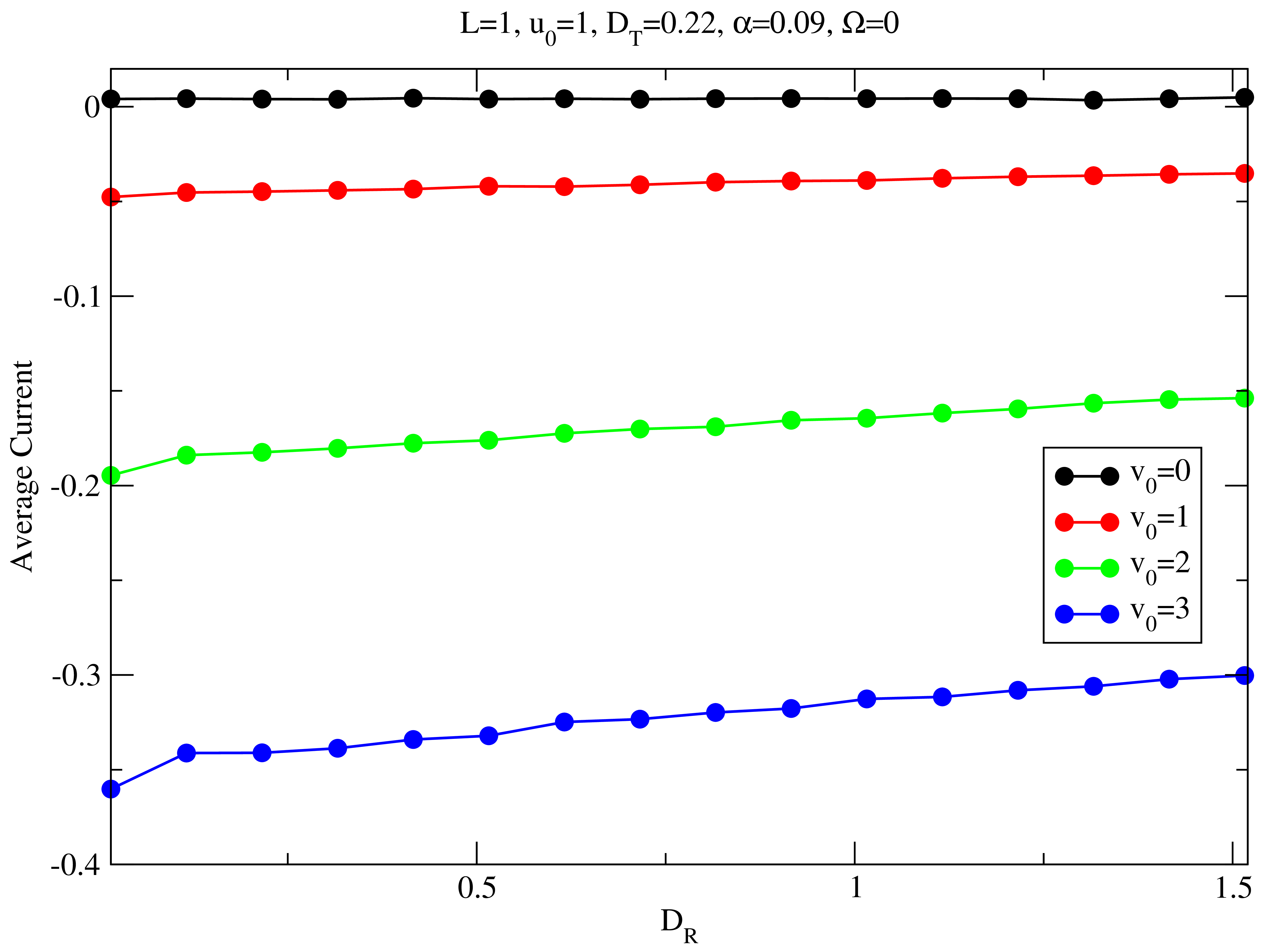}
  \includegraphics[scale=0.045]{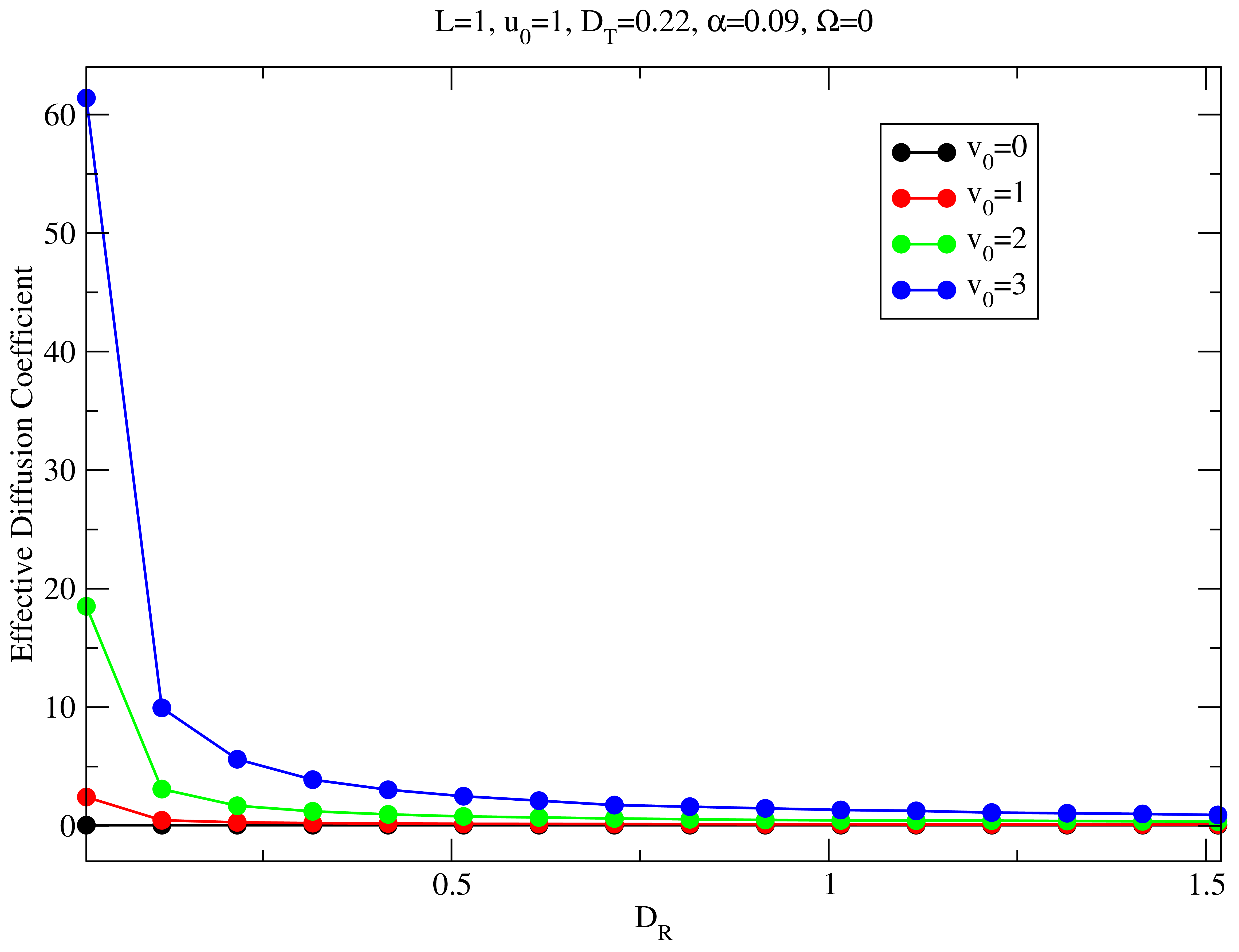}
  \caption{Average Current and Effective diffusion coefficient as a function of $D_R$ of an achiral particle for various values of $v_0$. The period and height of the potential is taken as unity, The translational diffusion coefficient is 0.22 and the value of asymmetric parameter is 0.09} 
 \end{figure}

\begin{figure}[H]
\includegraphics[scale=0.045]{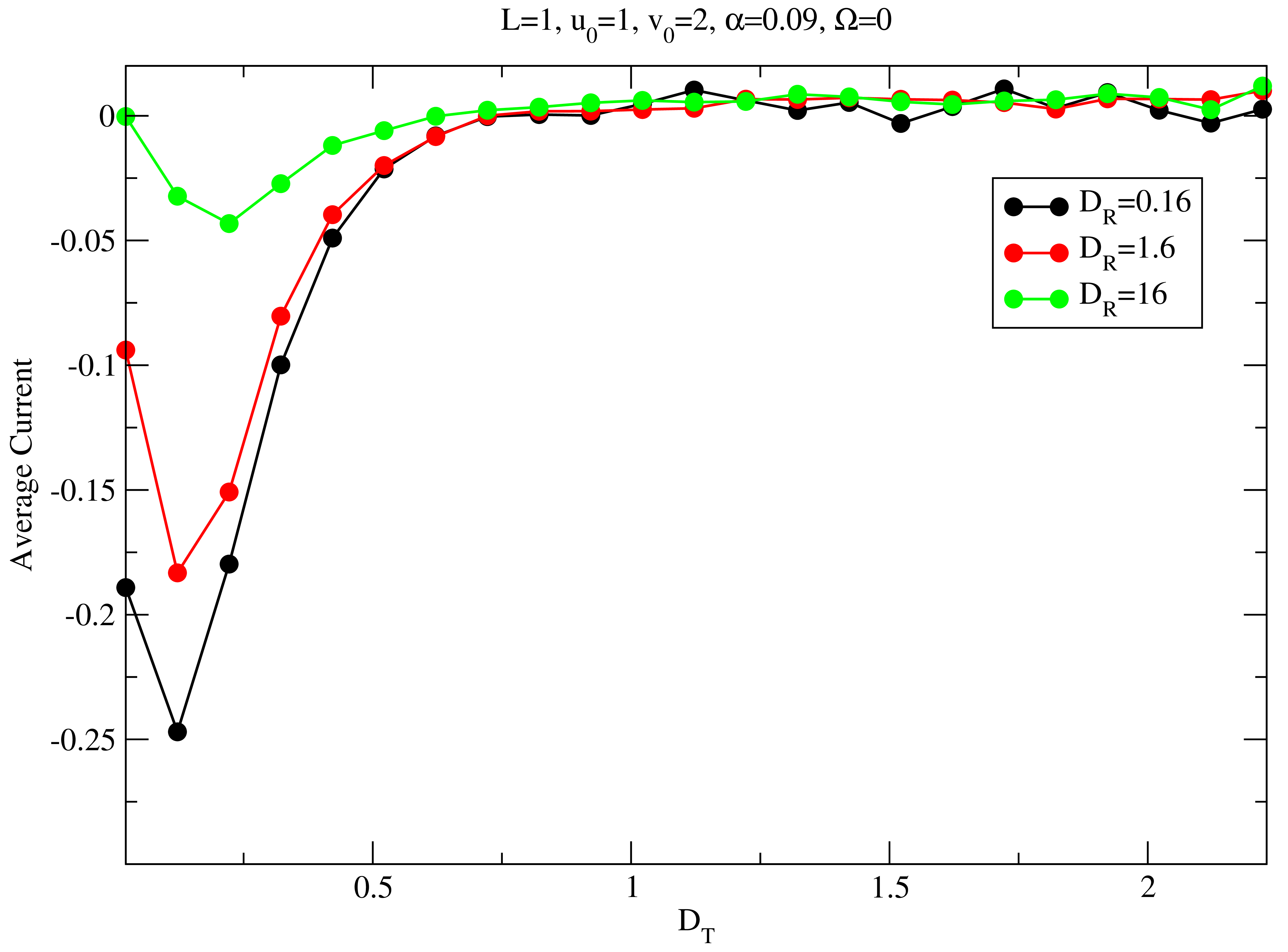}
\includegraphics[scale=0.045]{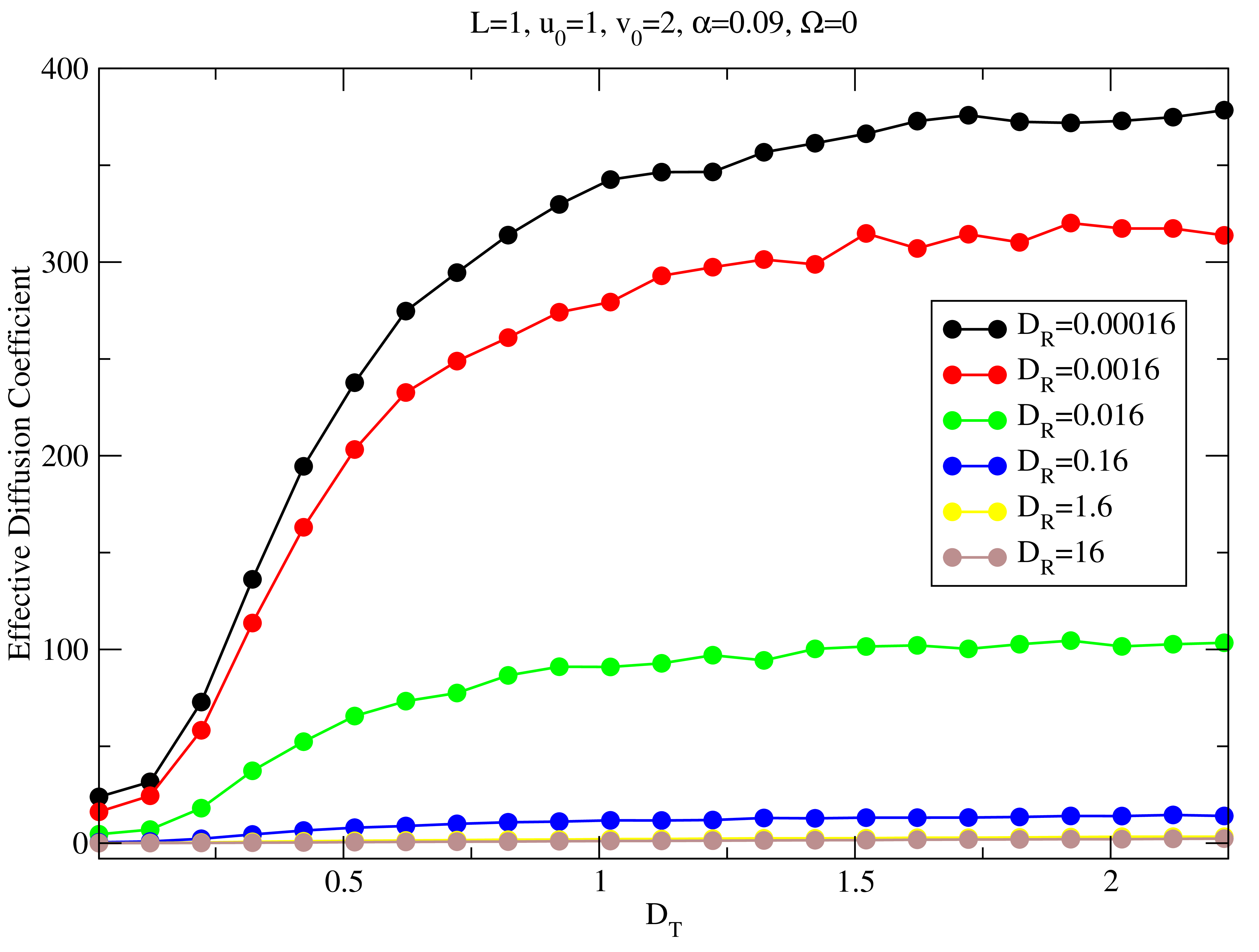}
\caption{Average Current and Effective diffusion coefficient as a function of $D_T$ of an achiral particle for various values of $D_R$. The period and height of the potential is taken as unity, The self-proppeled velocity is 2 and the value of asymmetric parameter is 0.09} 
  \end{figure}
\begin{figure}[H]
\includegraphics[scale=0.045]{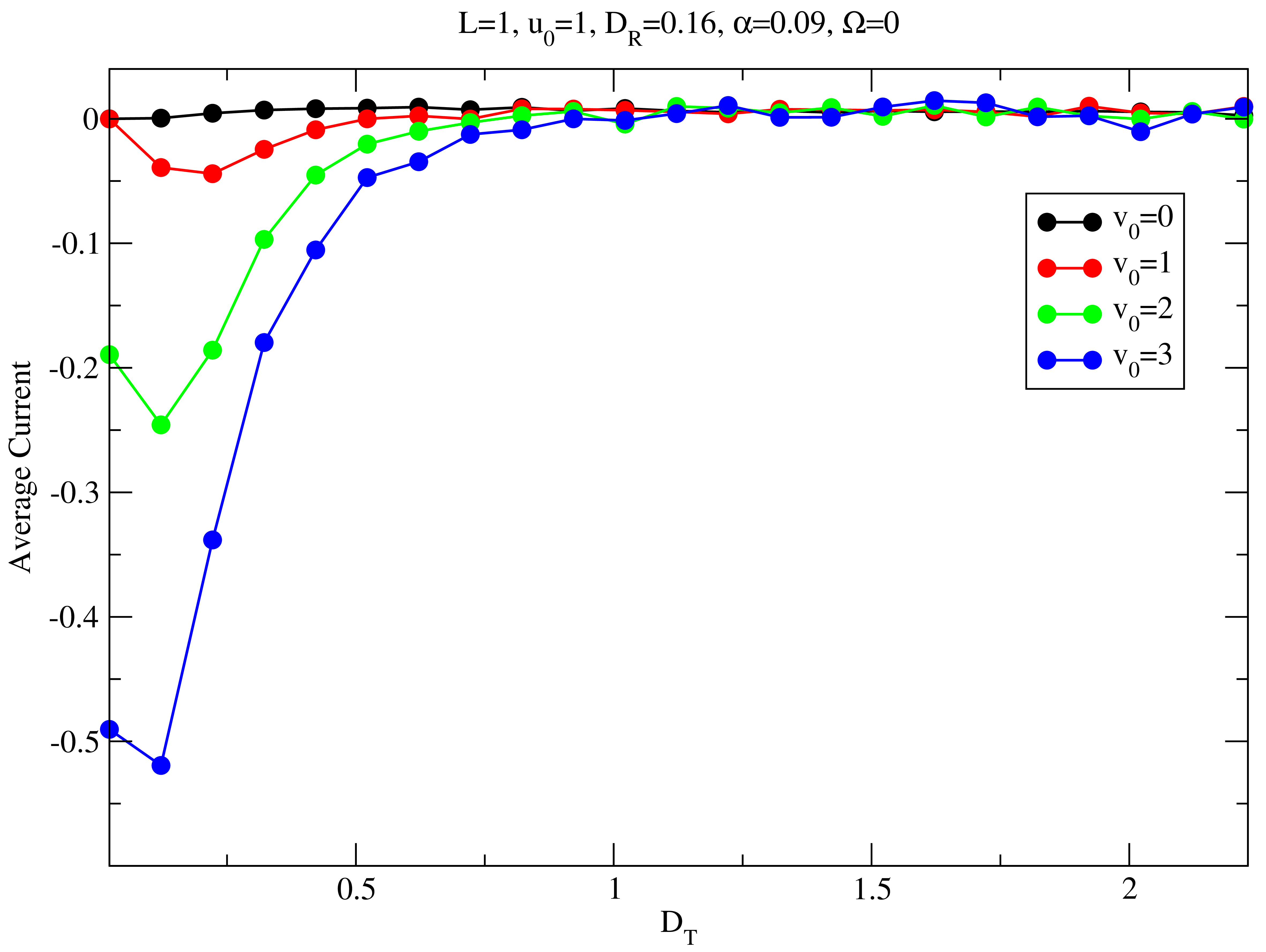}
\includegraphics[scale=0.045]{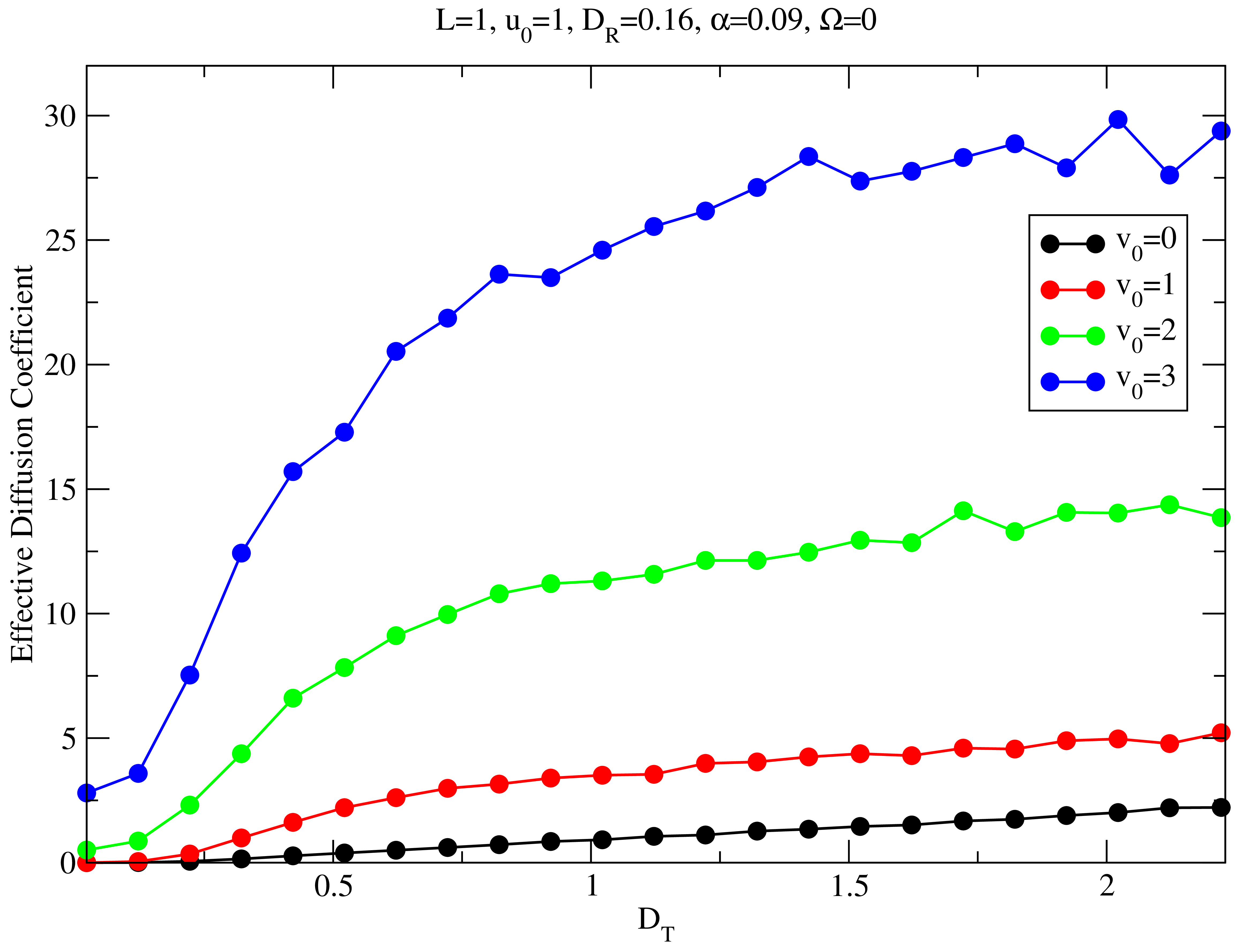}
\caption{Average Current and Effective diffusion coefficient as a function of $D_T$ for various values of $v_0$. The particle is achiral, the period and height of the potential is taken as unity, The rotational diffusion coefficient is 0.16 and the value of asymmetric parameter is 0.09} 

\end{figure}

\newpage
\begin{figure}[H]
\includegraphics[scale=0.045]{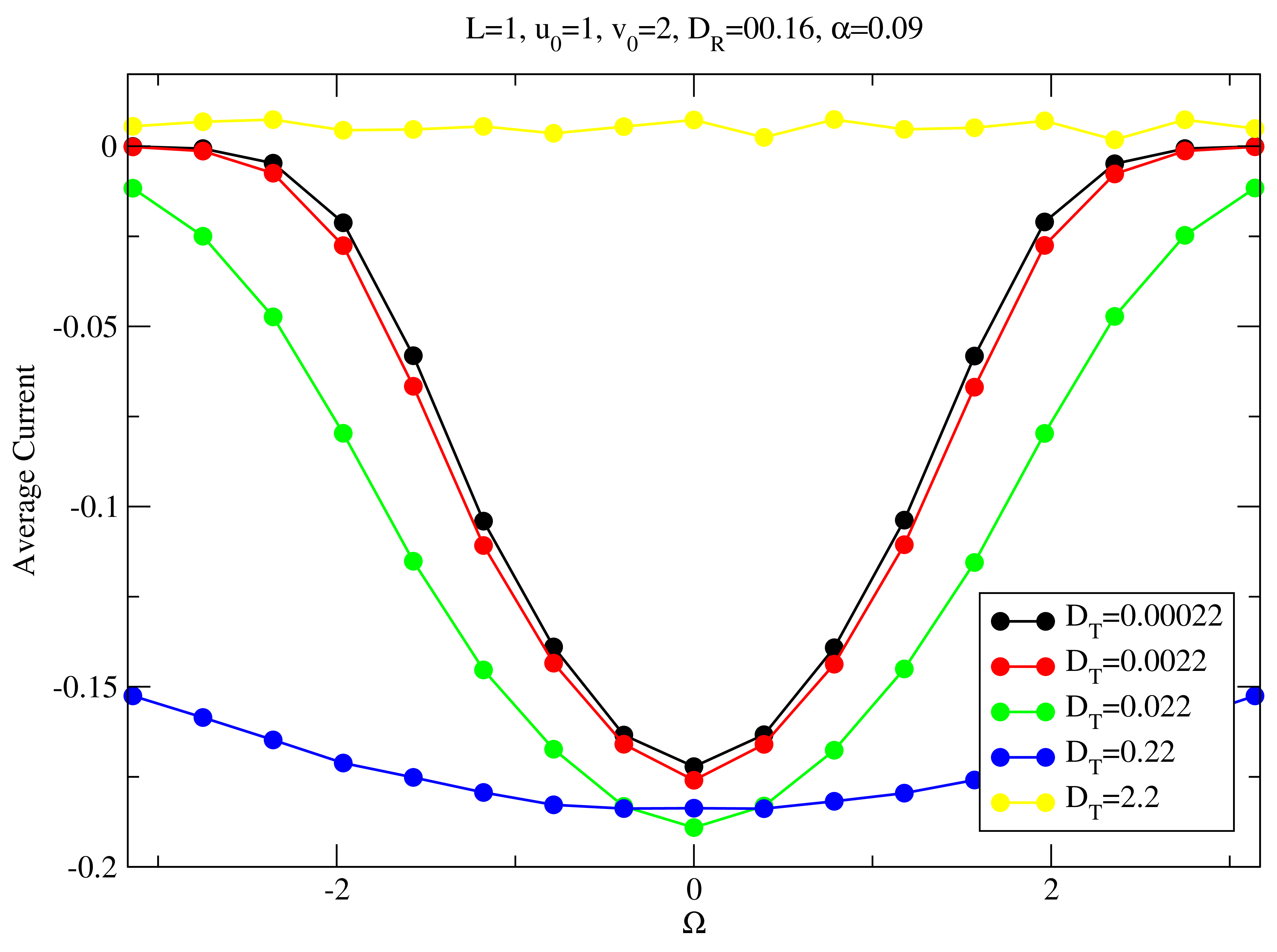}
\includegraphics[scale=0.045]{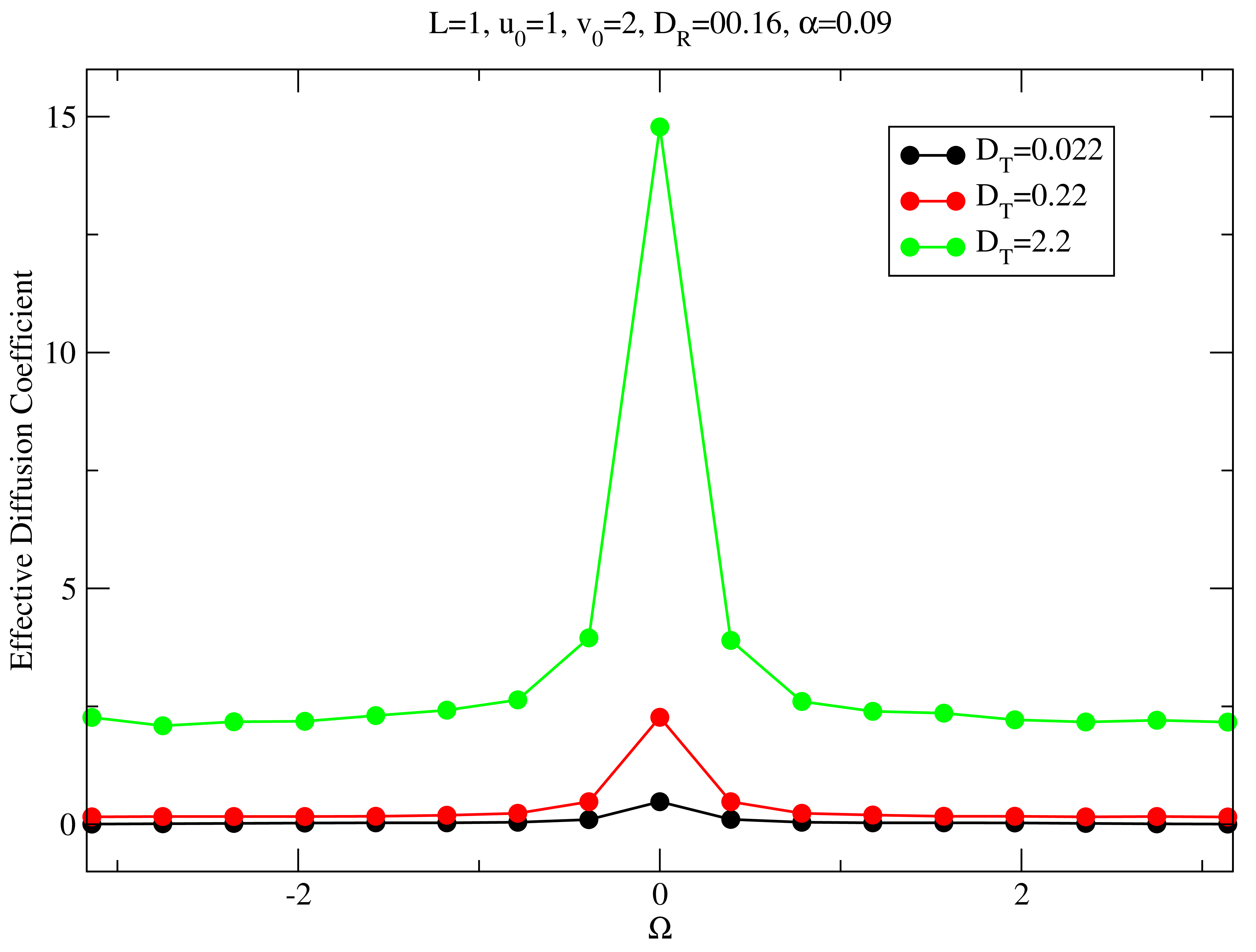}
\caption{Average Current and Effective diffusion coefficient as a function of $\Omega$ for various values of $D_T$. The period and height of the potential is taken as unity, The rotational diffusion coefficient is 0.16, the selfpropelled velocity is 2 and the value of asymmetric parameter is 0.09} 
  \end{figure}

\begin{figure}[H]
\includegraphics[scale=0.045]{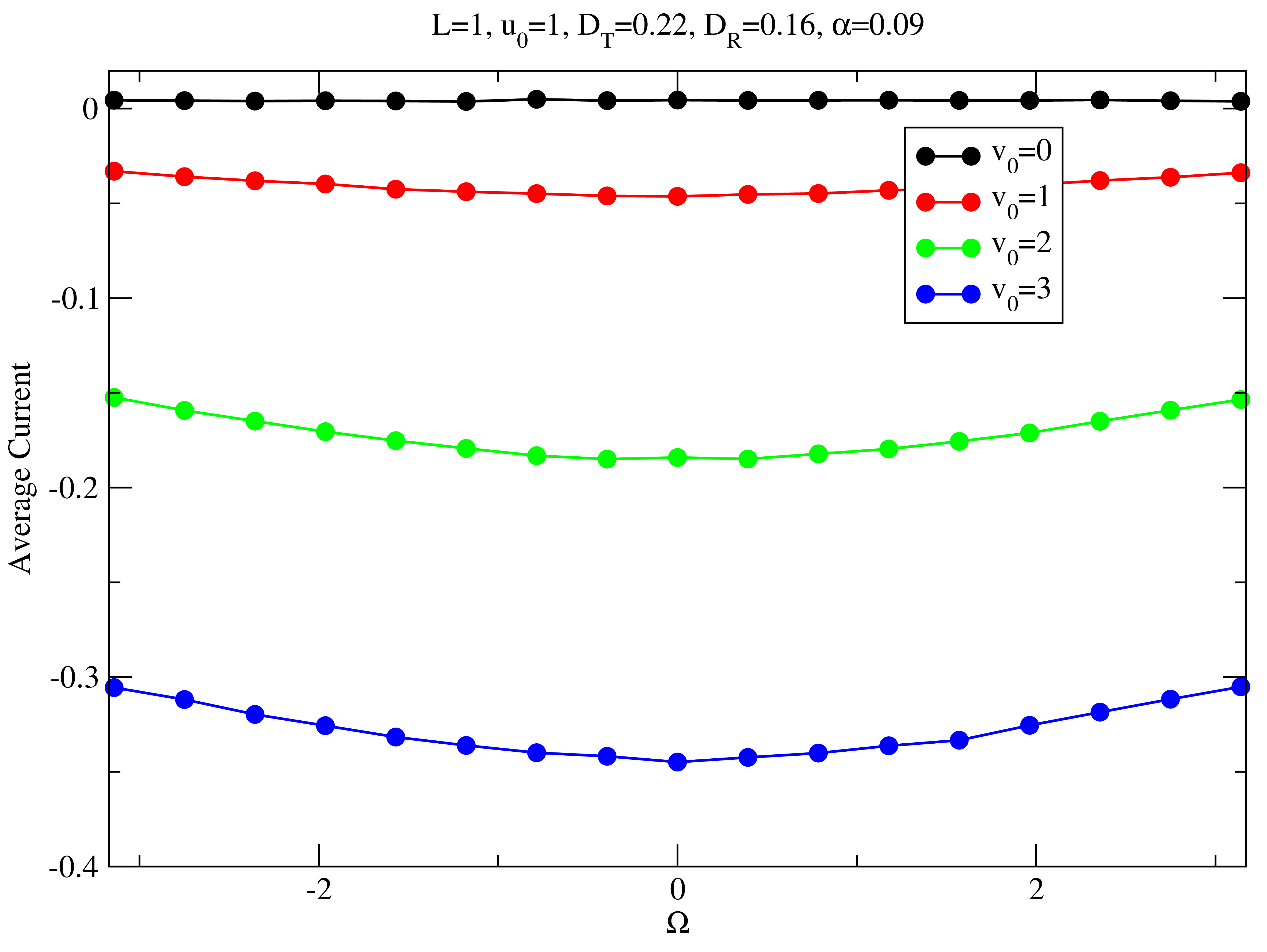}
\includegraphics[scale=0.045]{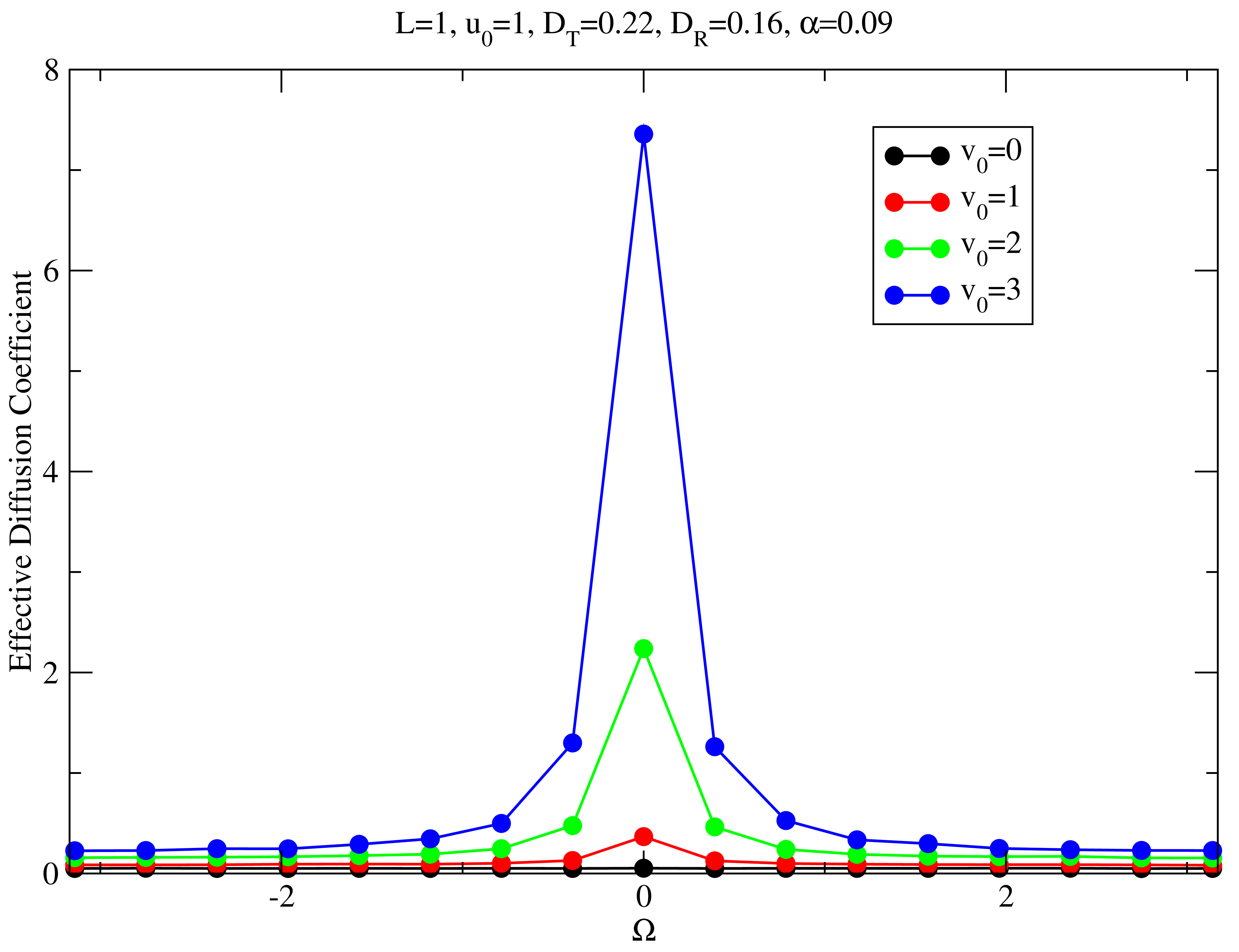}
\caption{Average Current and Effective diffusion coefficient as a function of $\Omega$ for various values of $v_0$. The period and height of the potential is taken as unity, The rotational diffusion coefficient is 0.16, translational diffusion coefficient is 0.22 and the value of asymmetric parameter is 0.09} 
\end{figure}

\begin{figure}[H]
\includegraphics[scale=0.045]{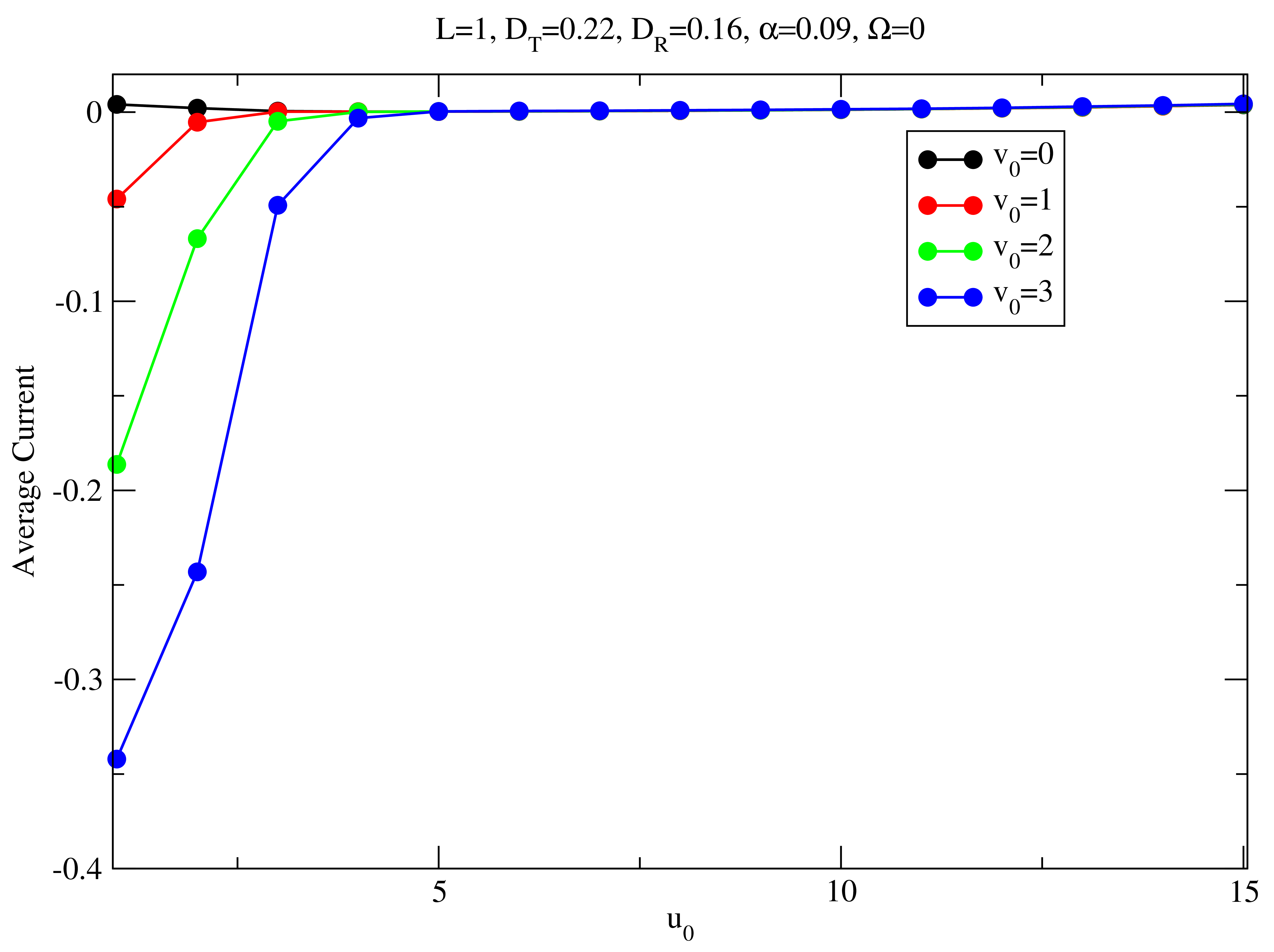}
\includegraphics[scale=0.045]{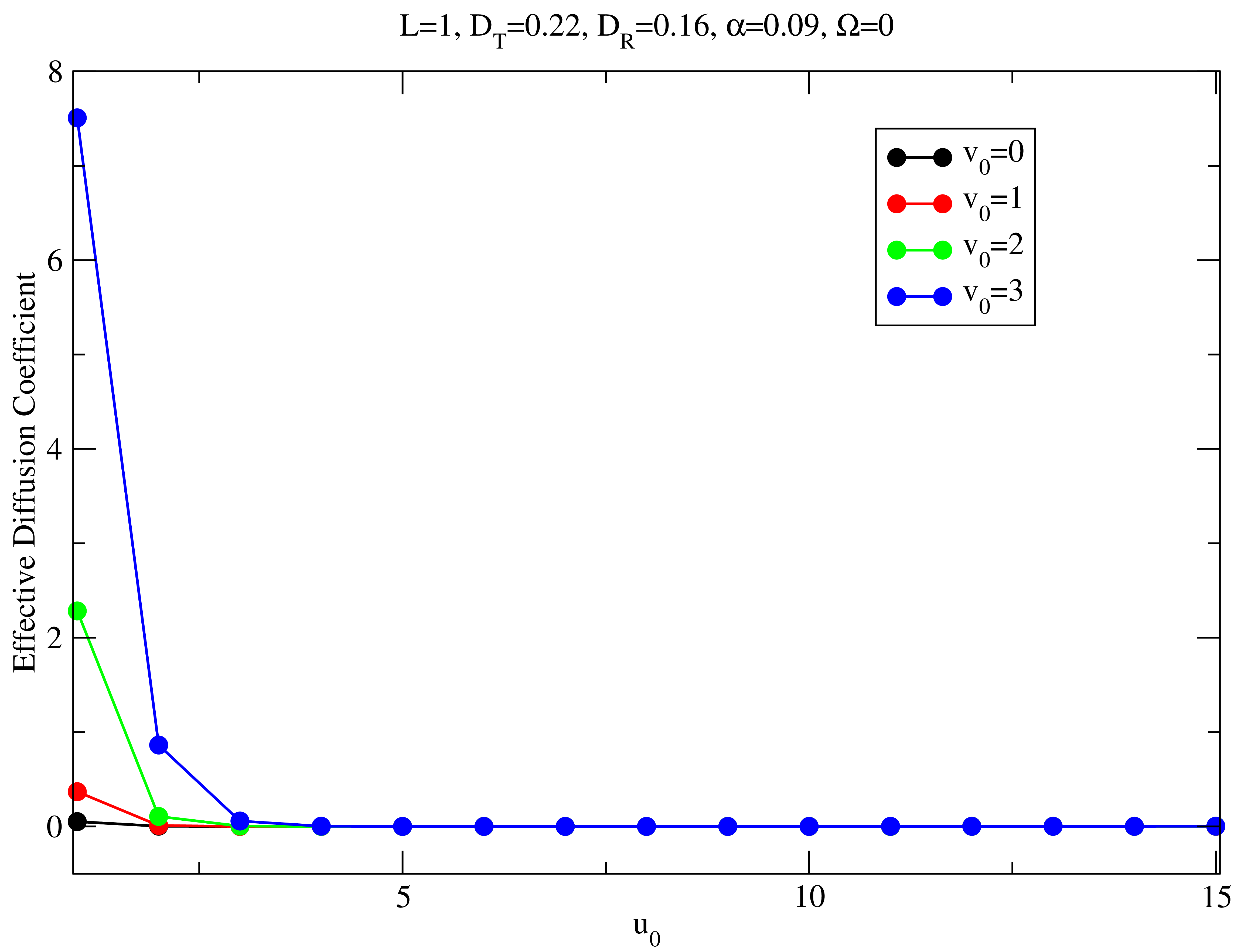}
\caption{Average Current and Effective diffusion coefficient as a function of $u_0$ for various values of $v_0$. The period of the potential is 1, The rotational diffusion coefficient is 0.16, the translational diffusion coefficient is 0.22  the value of asymmetric parameter is 0.09. The particle is achiral } 
  \end{figure}

\newpage
\begin{figure}[H]
\includegraphics[scale=0.045]{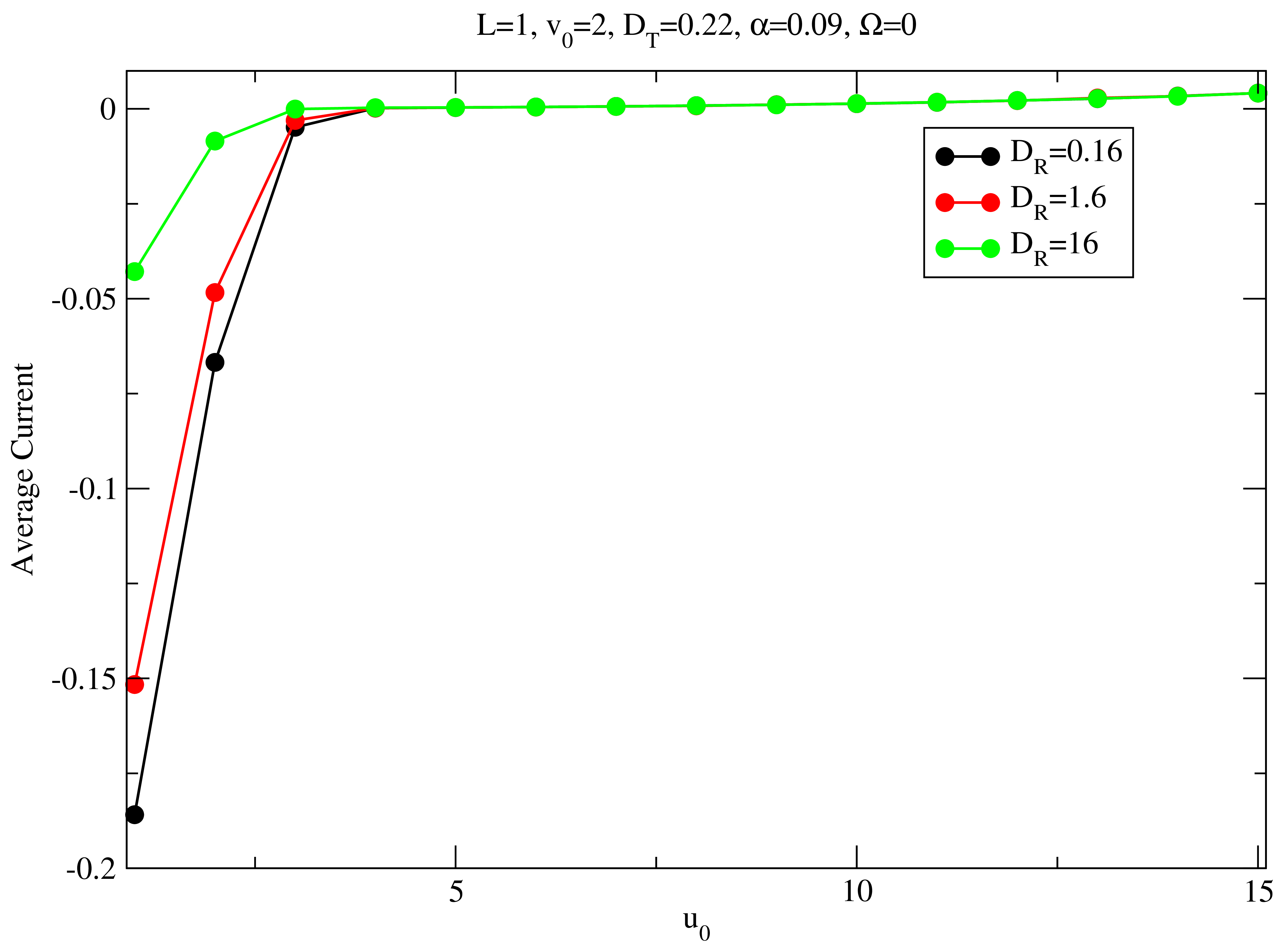}
\includegraphics[scale=0.045]{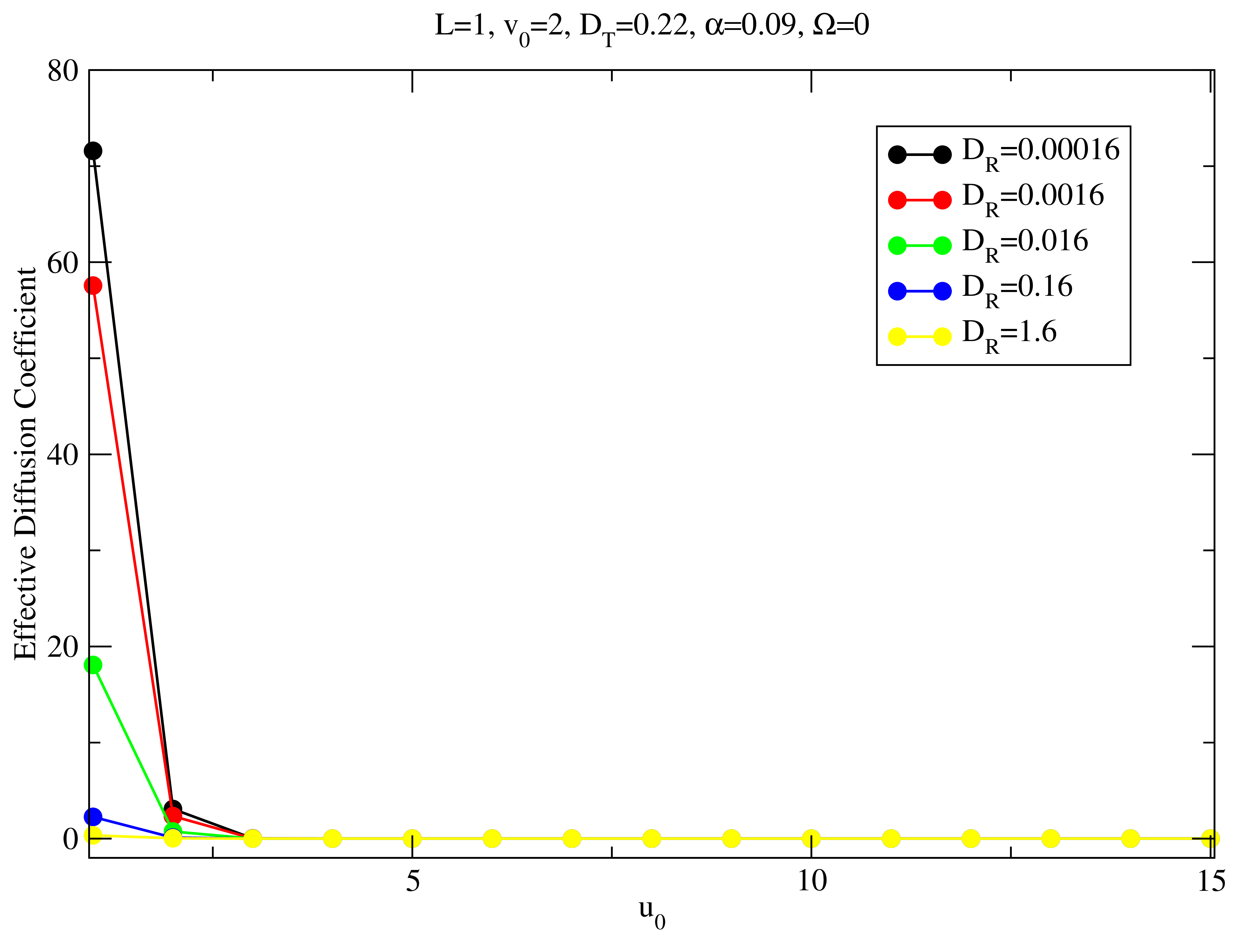}
\caption{Average Current and Effective diffusion coefficient as a function of $u_0$ for various values of $D_R$. The period of the potential is 1, the translational diffusion coefficient is 0.22, the self proppelled velocity is 2,  the value of asymmetric parameter is 0.09. The particle is achiral} 
\end{figure}

\begin{figure}[H]
\includegraphics[scale=0.045]{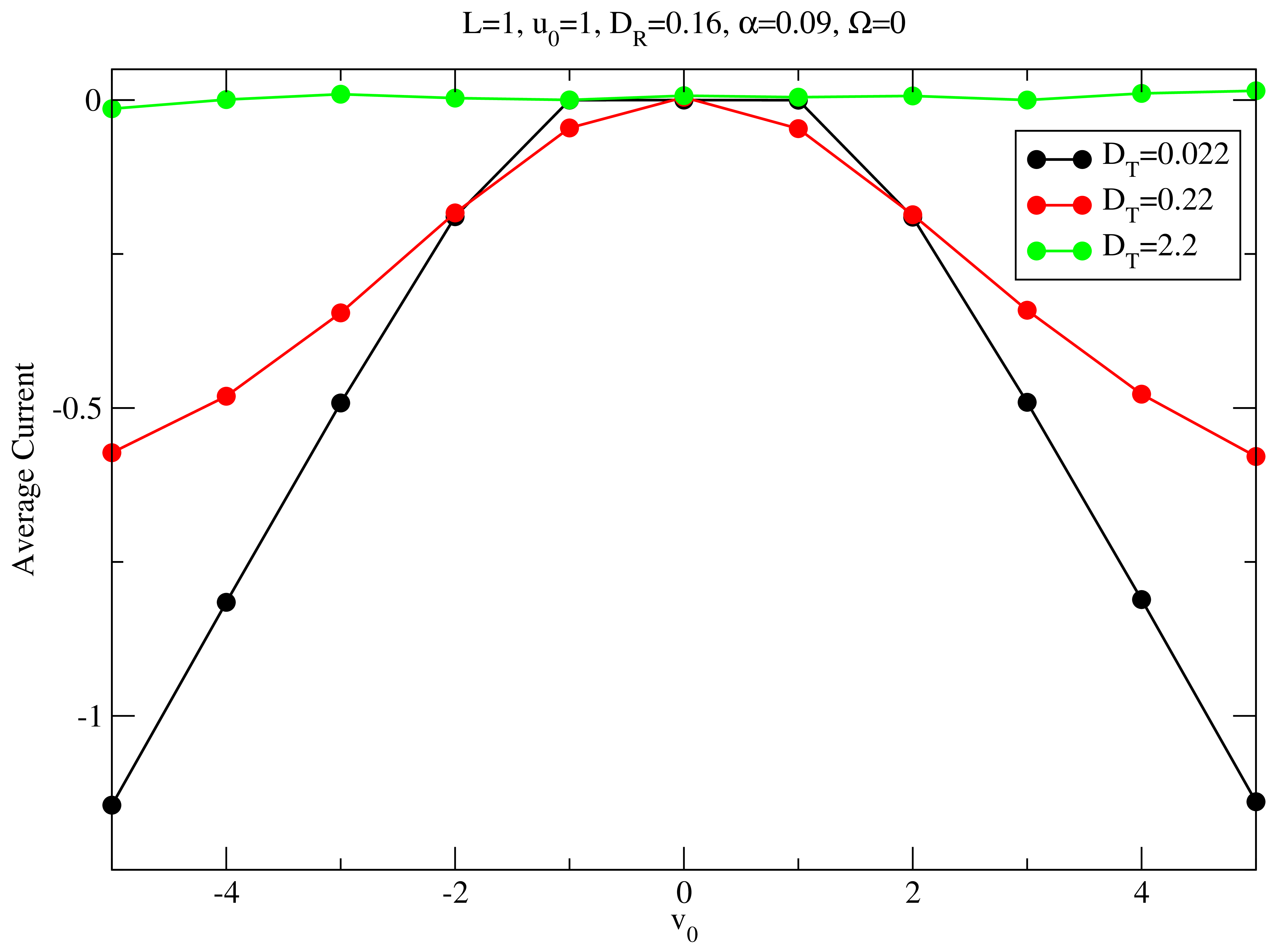}
\includegraphics[scale=0.045]{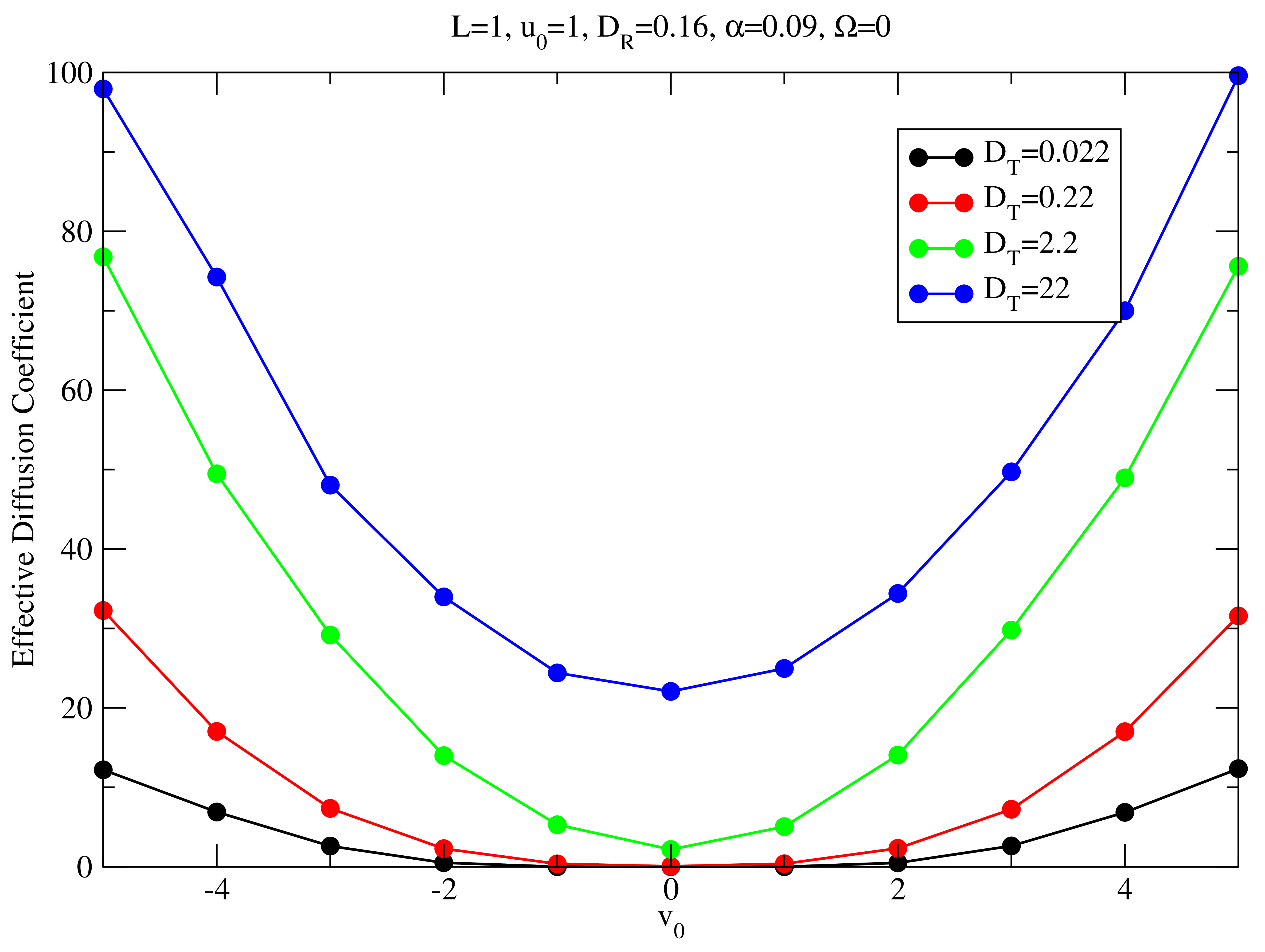}
\caption{Average Current and Effective diffusion coefficient as a function of $v_0$ of an achiral particle for various values of $D_T$. The period and height of the potential is taken as unity, The rotational diffusion coefficient is 0.16 and the value of asymmetric parameter is 0.09} 
\end{figure}
\begin{figure}[H]
\includegraphics[scale=0.045]{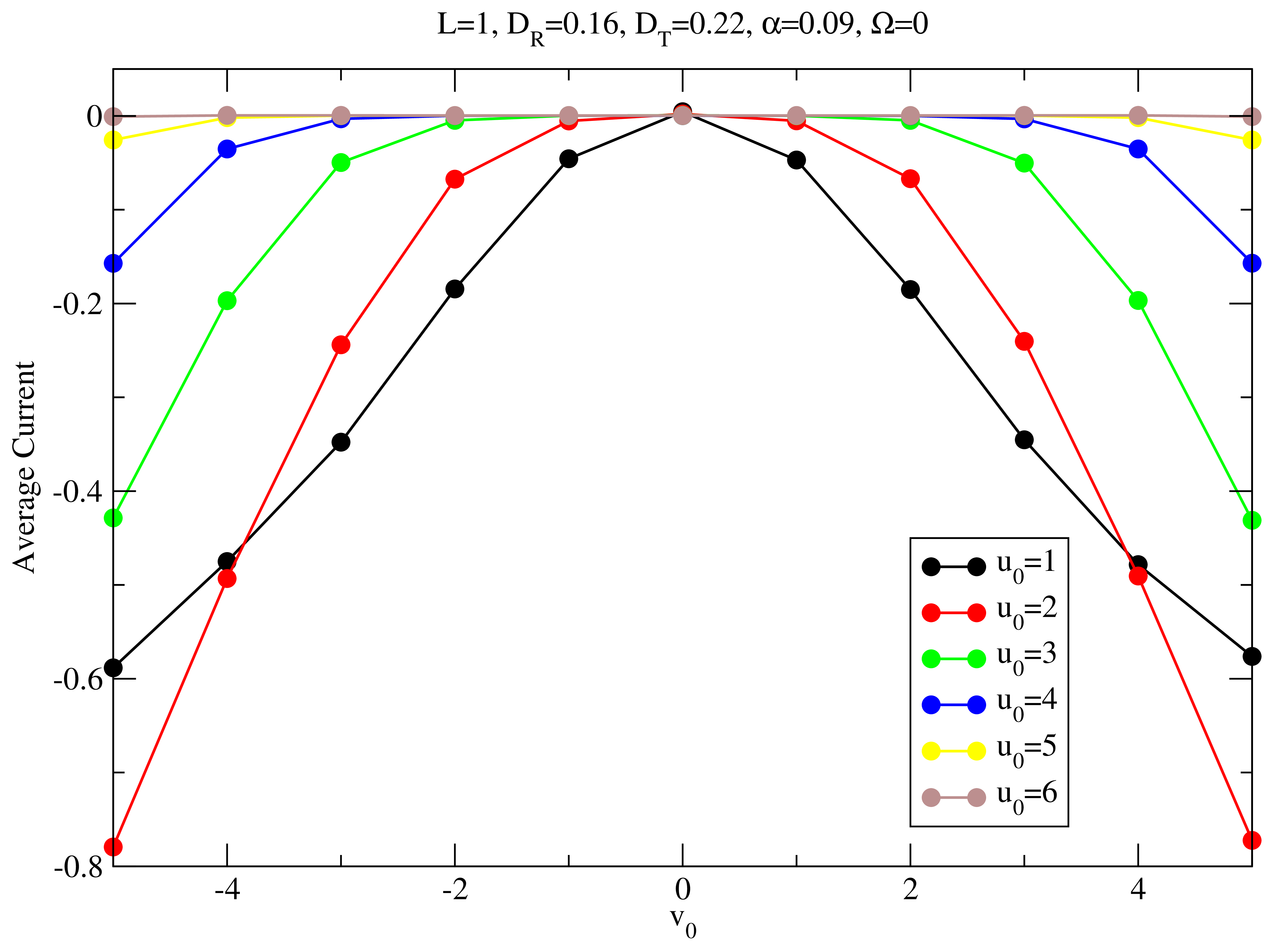}
\includegraphics[scale=0.045]{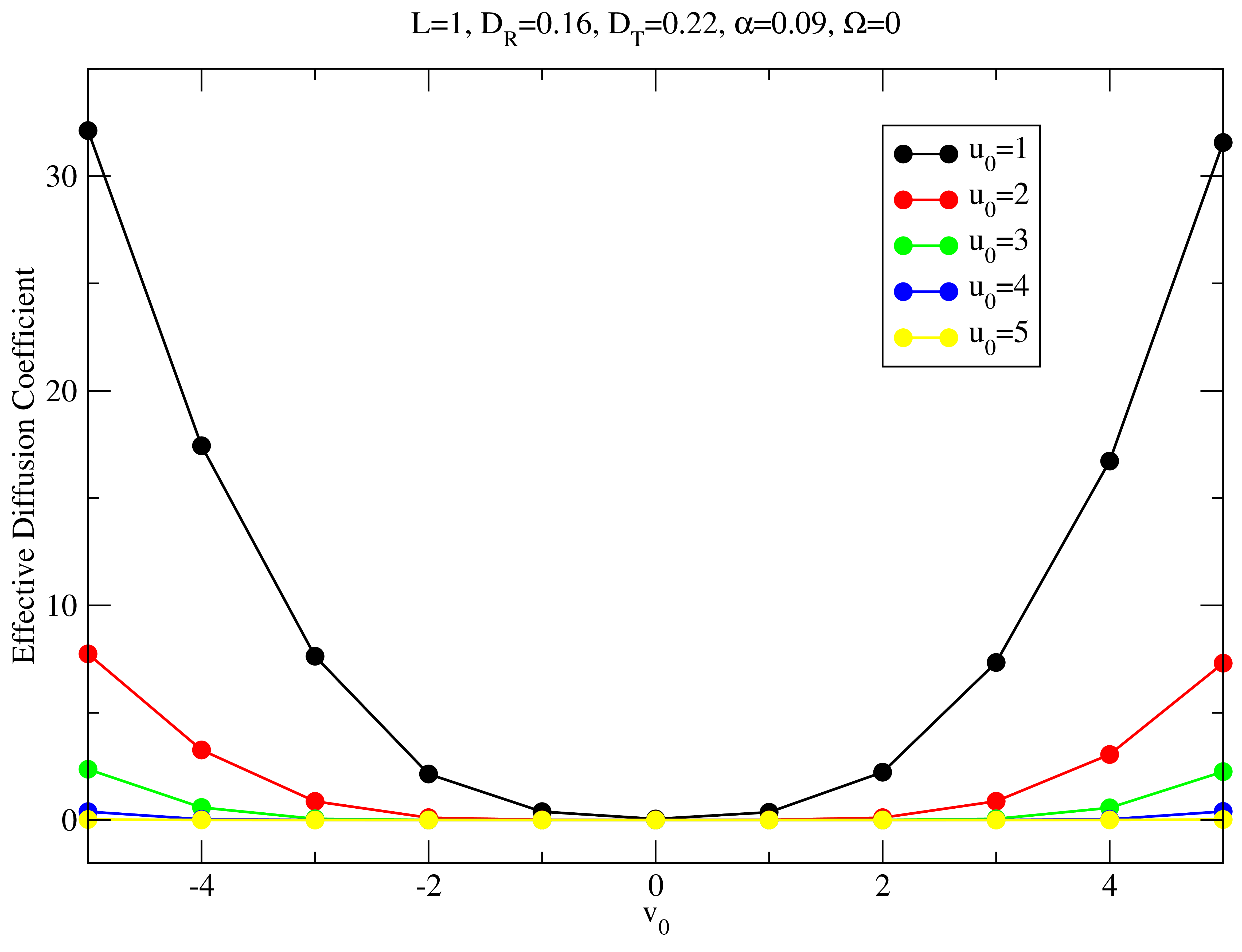}
\caption{Average Current and Effective diffusion coefficient as a function of $v_0$ for various values of $u_0$. The period is taken as unity, The rotational diffusion coefficient is 0.16, translational diffusion coefficient is 0.22 and the value of asymmetric parameter is 0.09. The particle is achiral.} 
\end{figure}
\newpage
\begin{multicols}{2}
\section{Conclusion}
The dynamics of the particle is studied from the behaviour of the current and the effective diffusion coefficient by changing various parameters that determines the motion of Active brownian particles. The current is generated only by active particles with self-proppelled velocity and only in asymmetric substrate. This two conditions satisfy the basic requirement for the system to come out of equilibrium. The direction of the self proppelled velocity(either positive or negative) and that of the angular velocity(clockwise or anticlockwise) does not alter the transport properties, since the rotational motion reorients the particle in each instict and cancels the effects. Another importent observation is chirality of the particle always reduces the current and confines the particle near its initial position. With the increasing value of chirality particle velocity decreases irrespective of its direction(clockwise or anti-clockwise). 
\bibliographystyle{ieeetr}
\bibliography{References}
\end{multicols}
\end{document}